\documentclass[%
reprint,
nofootinbib,
amsmath,amssymb,
aps,
]{revtex4-2}
\usepackage[utf8]{inputenc}
\usepackage{natbib}
\usepackage[table,dvipsnames]{xcolor}

\usepackage{subcaption}
\usepackage{graphicx}

\usepackage{dcolumn}% Align table columns on decimal point
\usepackage{bm}% bold math
\usepackage{physics}
\usepackage[colorlinks=true,linkcolor=blue,citecolor=blue,urlcolor=
blue]{hyperref}
\usepackage{latexsym}

\usepackage{lipsum}
\usepackage{mwe}
\usepackage{multirow}

\usepackage{float}

\newcommand{\Tbh}{\thickbar{h}}

\newcommand{\munu}{\mu\nu}

\newcommand{\lrb}[1]{\left( #1 \right)}
\newcommand{\lrrb}[1]{\left[ #1 \right]}

\newcommand\thickbar[1]{\mathbf{\bar{\text{$#1$}}}}

\begin{document}
	
	\preprint{APS/123-QED}
	%\title{Does Electric Field affect the Flow Harmonics in Heavy-Ion Collisions?}
	\title{}
	\title{Study of Early time attractor with Longitudinal Forces with Bjorken Symmetry   }% Force line breaks with \\
	%\thanks{The Bhatnagar-Gross-Krook~(BGK) approximation~\cite{Bhatnagar:1954zz}}%
	\author{Reghukrishnan Gangadharan}
	\email{reghukrishnang@niser.ac.in}%
	\author{Ankit Kumar Panda}
	\email{ankitkumar.panda@niser.ac.in}
	\author{Victor Roy}%
	\email{victor@niser.ac.in}
	\affiliation{School of Physical Sciences, National Institute of Science Education and Research,
		An OCC of Homi Bhabha National Institute, Jatni-752050, India}%

	\date{\today}% It is always \today, today,
	
	\begin{abstract}
  We study the effect of external proper time-dependent longitudinal forces on the evolution of the distribution function using the Boltzmann Equation with a relaxation time collision kernel under Bjorken flow. We derive an exact solution and study the early time attractor behavior of different components of energy-momentum tensor under Bjorken symmetry. We show that the arbitrary initial data approaches the attractor solution but is influenced by the external force with slightly broken Bjorken symmetry. 
	\end{abstract}
	
	\maketitle
	\section{Introduction} 
In the realm of high-energy heavy-ion collisions, two massive nuclei collide at relativistic speeds and create quark-gluon plasma (QGP)\cite{STAR:2005gfr}. This novel state of matter is composed of deconfined quarks and gluons, almost freely traversing the nuclear volume. Owing to the initial velocity of the two colliding nuclei, initially, the hot and dense fireball expands primarily along the beam direction; the transverse expansion due to the hydrodynamic response of the system takes some time to catch up with the longitudinal expansion\cite{PhysRevD.46.229,PhysRevC.82.039903,Romatschke:2009im,Heinz:2013th}. Due to this initial asymmetric flow, the momentum anisotropy can be substantial. This may give rise to a situation where the system is far away from local thermal equilibrium. The momentum anisotropy is also supposed to be larger for smaller systems such as p-p, p-Au, d-Au, etc. However, relativistic hydrodynamics seems to work well even for these small systems \cite{Dusling:2015gta,Li:2017qvf}.

Hydrodynamics is a valid description of dynamics when a system under scrutiny has attained a state of near local thermal equilibrium (i.e., the ratio of the off-equilibrium part of the energy-momentum tensor to the equilibrium one is much smaller than one). This raises the question of how the hydrodynamics can be a valid description for the QGP evolution at small time scales of $\sim$ $0.1$ to $0.2$ fm/c (at LHC) or for smaller systems where it is supposed to be highly anisotropic. Hydrodynamic attractors were proposed \cite{Heller:2015dha} as a possible solution to why hydrodynamics effectively describe QGP evolution even for systems far from local thermal equilibrium. For systems with certain symmetries- conformal and Bjorken/Gubser, it was observed that the non-hydrodynamic modes decay exponentially, and the system relaxes to an attractor solution regardless of the initial conditions (near or far away from local thermal equilibrium).
It was speculated then that the validity of hydrodynamics could be expanded beyond the limit of small gradients as previously held. Several subsequent studies ~\cite{Romatschke:2017acs,Jaiswal:2019cju,Chattopadhyay:2021ive,Jaiswal:2022udf,Jankowski:2023fdz,Kamata:2022jrc,Dash:2020zqx,Chattopadhyay:2019jqj,Giacalone:2019ldn,Blaizot:2019scw,Strickland:2018ayk,Blaizot:2017ucy,Behtash:2017wqg}, showed that attractor solutions could be observed in less symmetric systems albeit in a more general sense of an attractor.   However, more recent non-conformal studies  \cite{Chattopadhyay:2021ive, Jaiswal:2022udf,Alalawi:2022pmg} using exact solutions to the Boltzmann equation pointed out that attractor behavior is not as general as previously thought as it was absent in several hydrodynamic variables when conformal symmetry was broken and non-conformal systems only show an early-time attractor for the case of scaled longitudinal pressure $P_L/P$. 
 
Given that a universal attractor is absent in non-conformal systems, one could ask whether attractors still exist when other symmetries are broken. One method to explore this is to introduce an external force. Due to the relativistic speed of the charged protons inside the colliding nuclei, there exists a strong transient electromagnetic field in the initial stage of heavy-ion collisions.
%Additionally, during the initial stages of heavy ion collisions, alongside QGP production, formidable transient electromagnetic fields arise. 
These fields are estimated to possess strengths on the order of $10^{18}$ to $10^{19}$ G at the apex of RHIC or LHC energies~\cite{Gursoy:2014aka,Voronyuk:2011jd,Deng:2012pc,Zhao:2019crj,Kharzeev:2007jp}. Advancements have recently been made across various fronts, encompassing theoretical~\cite{Denicol:2018rbw,Denicol:2019iyh,Panda:2020zhr,Panda:2021pvq,Dash:2023kvr}, phenomenological~\cite{Alam:2021hje,Panda:2023akn}, and experimental domains~\cite{STAR:2023jdd,Das:2022lqh}, all in pursuit of detecting the telltale signals of these potent electromagnetic fields. Given that early-time attractors explore the journey toward equilibrium from highly non-equilibrated states, it is both logical and imperative to investigate their behavior in the presence of electromagnetic fields. In this study, we intend to explore the evolution of a far-from-equilibrium system under an external force arising due to the interaction of electrically conducting QGP fluid and the initial strong EM fields. 

% Given that a universal attractor is absent in non-conformal systems, one could ask whether attractors still exist when other symmetries are broken.  One method to explore this is to introduce an external force. Due to the relativistic speed of the charged protons inside the colliding nuclei, there exists a strong transient electromagnetic field in the initial stage of heavy-ion collisions.
% %Additionally, during the initial stages of heavy ion collisions, alongside QGP production, formidable transient electromagnetic fields arise. 
% These fields are estimated to possess strengths on the order of $10^{18}$ to $10^{19}$ G at the apex of RHIC or LHC energies~\cite{Gursoy:2014aka,Voronyuk:2011jd,Deng:2012pc,Zhao:2019crj,Kharzeev:2007jp}. Advancements have recently been made across various fronts, encompassing theoretical~\cite{Denicol:2018rbw,Denicol:2019iyh,Panda:2020zhr,Panda:2021pvq,Dash:2023kvr}, phenomenological~\cite{Alam:2021hje,Panda:2023akn}, and experimental domains~\cite{STAR:2023jdd}, all in pursuit of detecting the telltale signals of these potent electromagnetic fields. Given that early-time attractors explore the journey toward equilibrium from highly non-equilibrated states, it is both logical and imperative to investigate their behaviour in the presence of electromagnetic fields. In this study, we intend to explore the evolution of a far-from-equilibrium system under the presence of an external force arising due to the interaction of electrically conducting QGP fluid and the initial strong EM fields.

We use the RTA-Boltzmann equation to study the effects of external forces in a $0+1$ dimensional setting under approximate Bjorken symmetry. Specifically, we subject the system to a longitudinal force and explore how the evolution of the system varies for varying field strengths and initial anisotropies. We check for the presence of attractor behavior in hydrodynamic variables for the non-conformal systems and if they persist in the conformal limit. For our calculations, we will be using natural units, $\bar{h} = k_{B}= c = \epsilon_0 =\mu_0 = 1$, and the metric signature is mostly negative $g_{\munu} = diag(+,-,-,-)$.

Our paper is structured as follows. In Sec.\eqref{sec:Boltzmanneqn}, we discuss the relativistic Boltzmann equation with the RTA collision kernel. Then, in the next section, Sec.\eqref{sec:bjorken}, we discuss symmetries and the Bjorken flow. In the next section, Sec.\eqref{sec:solu}, we talk about the formal solution of the Boltzmann equation with external forces. We also cover the discussions on the initial distribution function and how to describe these external forces in Sec.\eqref{Sec:InitCond}. Finally, we present our findings for different situations involving external forces and summarize our findings in Sec.\eqref{sec:results},\eqref{sec:summary} respectively.

 \section{The Boltzmann Equation} {\label{sec:Boltzmanneqn}}
 The Boltzmann equation describes the evolution of the single-particle phase space distribution function $f(x,p)$ of a statistical system. The expression of the Boltzmann equation in the presence of external forces can be written as \cite{Cercignani:2002rbe}  
 \begin{align}
     p^{\mu}\pdv{f}{x^{\mu}} + mK^{\mu}\pdv{f}{p^{\mu}} - \Gamma^{\sigma}_{\munu}p^{\mu}p^{\nu}\pdv{f}{p^{\sigma}} = C(f)\,,
 \end{align}
here $K^{\mu}$ is a 4-force and $\Gamma^{\rho}_{\mu\nu}$ is the Christoffel symbol. The momenta $p^{\mu}$ satisfy the onshell relation $g_{\munu}p^{\mu}p^{\nu} = m^2$ and the 4-force satisfy $g_{\munu}p^{\mu}K^{\mu}/p^{0} = 0$. 
As the general collision kernel is quite complex, we use Anderson-Witting relaxation time approximation(RTA) \cite{Cercignani:2002rbe}
\begin{align}
    C(f) = -u\cdot p\frac{(f-f_{eq})}{\tau_R}.
\end{align}
Here, $f_{eq} \sim e^{-(u \cdot p)/T}$ represents the local equilibrium distribution function.  $u \cdot p = g_{\mu\nu}u^{\nu}p^{\mu}$ represents the inner product. $T$ is the local temperature and $u^{\mu}$ is the fluid 4-velocity ($g_{\mu\nu}u^{\mu}u^{\nu} = 1$). We consider the relaxation time $\tau_R$ is a function of temperature but not of momentum. 

The energy-momentum tensor $T^{\munu}(x)$ is obtained from the distribution function by taking its second moment
\begin{align}\label{Eq:EMTDef}
    T^{\munu} = \int d{\Xi} p^{\mu}p^{\nu}f ,
\end{align}
where the integration measure is given by $d\Xi = \frac{\sqrt{-g}d^3p}{(2\pi)^3p^{0}}$. 
The energy-momentum conservation $\nabla_{\mu}T^{\munu} = 0 $ is ensured for the RTA kernel by imposing the Landau matching condition
\begin{align}\label{Eq:LMC}
    \int d{\Xi} (u_{\mu}p^{\mu})^2f = \int d{\Xi} (u_{\mu}p^{\mu})^2 f_{eq},
\end{align}
where $\nabla_{\mu}$ is the covariant derivative. Here the fluid four-velocity $u^{\mu}$ is defined as the Eigenvector of $T^{\mu\nu}$ with energy density $\epsilon$ as the Eigenvalue
\begin{align}\label{Eq:LFrame}
    T^{\munu}u_{\nu} = \epsilon u^{\mu}.
\end{align} 
%The imposition of Landau frame matching \eqref{Eq:LFrame} implies the energy-momentum conservation
%\begin{equation}
%    \nabla_{\mu}T^{\munu} = 0 \,,
%\end{equation}. 

%\subsection{Hydrodynamic variables}

 \section{Symmetries and Bjorken Flow} {\label{sec:bjorken}}

 The symmetries of the system set restrictions on the flow profile of the fluid. A Bjorken system is characterized by translation, rotational symmetry in the transverse plane ($x-y$), boost invariance along the longitudinal direction $z$, and a reflection symmetry($z \rightarrow -z$). The symmetries completely determine the Bjorken flow profile,
 \begin{align}
     u^{\mu} = \left(\frac{t}{\sqrt{t^2-z^2}},0,0,\frac{z}{\sqrt{t^2-z^2}}\right) \,.
 \end{align}
We use Milne coordinate system $(\tau, x, y ,\eta)$, where 
 \begin{align}
     \tau = \sqrt{t^2 -z^2},\\
     \eta = \tanh^{-1}{\frac{z}{t}},
 \end{align}
 where $\tau$ is the longitudinal proper time, $\eta$ the space-time rapidity. The Milne metric is given by $g_{\munu} = diag(1,-1,-1,-\tau^2)$ with the non zero Christoffels symbols
 \begin{align}
\Gamma^{\tau}_{\eta\eta} =  \frac{1}{\tau}  &,& \Gamma^{\eta}_{\tau\eta} =  \tau.
%\Gamma^{r}_{\phi\phi} =  \frac{1}{r}  && \Gamma^{\phi}_{r\phi} =  -r 
 \end{align}
 
 %The symmetry constraints significantly reduce the dimensional dependence of the problem from $3+1$ D to $0+1$ D. In the Minlne coordinates, the reduction in coordinate dependence becomes manifest as the flow profile is such that a transformation to the Minlne coordinates is equivalent to a local Lorentz transformation to the local fluid rest frame. 
 The flow profile in the Milne coordinate with Bjorken symmetry ($\eta$ invariance)  becomes $u^{\mu} = (1,0,0,0)$.  Rotational and translational symmetries in the transverse plane imply that the distribution function cannot depend on $x$ and $y$. The spatial dependence then reduces to only that of the proper time $\tau$ due to boost invariance and reflection symmetry.  We use transverse momentum $p_T = \sqrt{p_x^2 + p_y^2}$ and $w \equiv p_{\eta} = \tau p^{z} =tp^{z} - zp^{t}$ in the longitudinal direction. The distribution function is then only dependent on three variables, one space, and two momenta and the Boltzmann Equation in Milne coordinates takes the simple form 
 \begin{align}
     \pdv{f(\tau,p_T,w)}{\tau} = -\frac{(f-f_{eq})}{\tau_R}\,.
 \end{align}
 A formal solution can be found for this equation in \cite{Baym:1984np,Florkowski:2013lza,Florkowski:2013lya,Florkowski:2014sfa} which is given as:
 \begin{align}
    f(\tau,w,p_T) =& D(\tau,\tau_0)f_0(w,p_T) \\&+ \int_{\tau_0}^\tau \frac{d\tau'}{\tau_{R}(\tau')}D(\tau,\tau')f_{eq}(\tau',w,p_T)\,,
\end{align}
where $D(\tau_2,\tau_1)$ the damping function defined as
\begin{equation}
    D(\tau_2,\tau_1) = \exp\left[ -\int_{\tau_1}^{\tau_2}\frac{d\tau''}{\tau_R(\tau'')}\right] \,, 
\end{equation}
and $f_0 = f(\tau=\tau_0, w, p_T)$.

\subsection{RTA Boltzmann Equation under Longitudinal Force}
 The introduction of an external force is bound to break one of the four symmetries. This means that Bjorken flow could be modified in the presence of force. In order to approximately retain Bjorken flow, the simplest force one can then introduce is one that only depends on the proper time $\tau$ and is along the $z$ direction. The introduction of this force will break reflection symmetry about $\eta=0$. Considering the force to be weak enough so that the change in fluid velocity ($\delta u^{\mu}$) is small and the new velocity is written as
 \begin{align}
     u^{\mu} = u^{\mu}_B + \delta u^{\mu}.
 \end{align}
  %Here we make the crucial assumption that the flow profile is approximately Bjorken. 
  This approximation is valid as long as the correction to the fluid variables due to the force is smaller than the equilibrium values. Assuming Bjorken flow, the Boltzmann equation with force along $z$ direction can then be written as (see App.\eqref{sec:appA} for details)
 \begin{align}
 \label{eq:BoltzForce}
     \lrb{  \mathcal{F}(\tau)\tau\pdv{}{w} + \pdv{}{\tau} }f = -\frac{\lrb{f-f_{eq}}}{\tau_R} \, .
 \end{align}
 %\frac{\alpha}{\tau_m}  F\lrb{\tau/\tau_m}

\section{Solution under external force} {\label{sec:solu}}
Here we assume $\mathcal{F}(\tau )\neq 0 \,$ for any value of $\tau_i < \tau < \tau_f$ (where $\tau_i$ and $\tau_f$ are initial and final time).For convenience we introduce a dimensionless function F($\tau$) and express $\mathcal{F}=\frac{\alpha}{\tau_F}F(\tau)$. After a change of variables (see App.\eqref{Ap:ChofVar}), Eq.\eqref{eq:BoltzForce} takes the following form,
 \begin{align}  
    \pdv{f(r,s,p_T)}{s}  = -\frac{\tau_F}{\tau_R}\frac{\lrb{f-f_{eq}}}{F(\tau(s))\,\tau(s) \, } \,.
 \end{align}
Where we have defined a new variable $s$.
 \begin{align}
     s(\tau) &= \int_{0}^{\tau}F\lrb{\tau' }\frac{\tau'}{\tau_F}  d\tau' \,,    
 \end{align}
and used the coordinate transformations $(w,\tau) \to (r,s)$,
 \begin{align}
     r &= w - \alpha s \,,\\
     s &= s \,.
 \end{align}
Here $\alpha$ and $\tau_F$ are constants with dimensions of momentum and time respectively so that $\alpha/\tau_F$ has dimensions of force. 
Then we can write a solution 
 \begin{align}
 \label{Eq:fneq}
    f(s,r,p_{T}) = D(s,s_0)&f_{0}(r,p_{T})  \nonumber\\
                    &+ \int_{s_0}^{s}\frac{ds'}{\tau_R(s')}D(s,s')f_{eq}(s',r,p_{T}),
\end{align}
where 
\begin{align}
    D(s_{1},s_{2}) = \exp\lrrb{-\int_{s_1}^{s_2}\frac{ds'}{F(\tau(s'))\,(\tau(s')/\tau_F)\,\tau_R(s')}}.
\end{align}
 The above expression for $D(s_{1},s_{2})$ gives the impression that it depends explicitly on the force. However, if one were to change the coordinates back from $s\to\tau$, we can see that $D(s_{1},s_{2})$ takes the form
	
\begin{align}\label{Eq:Damp}
    D(\tau_{1},\tau_{2}) = \exp\lrrb{-\int_{\tau_1}^{\tau_2}\frac{d\tau'}{\tau_R(\tau')}} \, .
\end{align}	

Therefore dependence of $D(\tau_{1},\tau_{2})$ on force is only implicitly through $\tau_R(\tau')$. 

\section{Initial Conditions}\label{Sec:InitCond}

The initial anisotropic distribution is chosen to be that of Romatschke-Strickland and form \cite{Romatschke:2003ms},
\begin{align}
    f_0 = \frac{2}{(2\pi)^3 N_0}\exp{-\frac{\sqrt{p_T^2 +(1+\xi_0)p_z^2+m^2}}{\Lambda_0}}.
\end{align}
The three initial free parameters $N_0$, $\xi_0$ and $\Lambda_0$  allow us to specify the initial fluid parameters $T$,$P_L$ and $P_T$.  Here $\xi_0$  corresponds to the longitudinal anisotropic parameter in momentum space, and $\Lambda_0$ is an energy scale, and in the limit  it reduces to local temperature.  The value of $\Lambda_0$ and $N_0$ is set such that the energy density matches the energy density of an equilibrium distribution with the initial temperature $T_0$. To study the effect of varying $\frac{m}{\Lambda_0}$  we can change $N_0$ while ensuring that the initial energy density and $\xi_0$ remain fixed. For the non-conformal case ($m/\Lambda_0 \neq 0$) we use the same set of initial parameters (Tab.\eqref{tab:Init_Pi_Jaiswal},App.\eqref{tab:Init_Jaiswal})  as 
 \cite{Jaiswal:2022mdk} for ease of comparison. 

 %\begin{table}
 %   \begin{tabular}{|c|c|c|c|c|c|c|c|}\hline
 %      No.&  0 & 1 & 2 & 3 & 4 & 5 & 6\\ \hline
 %     $(\Pi/P)_{0}$&  0 & -0.25 & -0.37 & 0 & 0 & 0.25 & -0.85\\ \hline
 %     $(\pi/P)_{0}$ &  -1& -1 & -1& 0.99& -1.8& 0 & 0 \\ \hline
%\end{tabular} 
%\caption{Various initial values of$(\Pi/P)_{0}$,$(\pi/P)_{0}$ and their number code}
%\end{table}\label{Tab:InitPi}

 The hydrodynamic variables like energy density~($\mathcal{E}$), longitudinal $P_L$ and transverse pressure $P_T$ in the presence of external forces can be obtained from Eq.\eqref{Eq:EMTDef} and Eq.\eqref{Eq:fneq} 
\begin{align} 
     {\mathcal{E}}(s) &= D(s,s_0)\frac{\Lambda_{0}^{4}}{4\pi N_0}{H}^{F}_{\epsilon}\lrrb{s_0,s,\xi_0,\frac{\alpha}{\Lambda_{0}},\frac{m}{\Lambda_{0}}} \nonumber\\
     &+ \int_{s_0}^{s}\frac{ds'}{\tau_R(s')}D(s',s)T^{4}(s'){H}^{F}_{\epsilon}\lrrb{s',s,\frac{\alpha}{T},\frac{m}{T}}\label{eq:E-Tau} \,,\\
     P_L(s)  &= D(s,s_0)\frac{\Lambda_{0}^{4}}{4\pi N_0}{H}^{F}_{L}\lrrb{s_0,s,\xi_0,\frac{\alpha}{\Lambda_{0}},\frac{m}{\Lambda_{0}}} \nonumber\\
     &+ \int_{s_0}^{s}\frac{ds'}{\tau_R(s')}D(s',s)T^{4}(s'){H}^{F}_{L}\lrrb{s',s,\frac{\alpha}{T},\frac{m}{T}}\label{eq:PL-Tau}\,, \\
     P_T(s) &= D(s,s_0)\frac{\Lambda_{0}^{4}}{4\pi N_0}{H}^{F}_{T}\lrrb{s_0,s,\xi_0,\frac{\alpha}{\Lambda_{0}},\frac{m}{\Lambda_{0}}} \nonumber\\
     &+ \int_{s_0}^{s}\frac{ds'}{\tau_R(s')}D(s',s)T^{4}(\tau'){H}^{F}_{T}\lrrb{s',s,\frac{\alpha}{T},\frac{m}{T}}\,,\label{eq:PT-Tau} 
\end{align}\label{Eq:IntEPT}
by evaluating the corresponding integrals App.[\ref{App:Int}]. Note that due to the breaking of reflection symmetry, the energy-momentum tensor is no longer diagonal. The non-diagonal term $T^{03}$ can be evaluated as
\begin{align}
   T^{03}  = \int_{s_0}^{s}\frac{ds'}{\tau_R(s')}D(s,s')T^{4}(s'){H}^{F}_{03}\lrrb{s',s,\frac{m}{T}}.\label{eq:T03-Tau}
\end{align}
We can use it to calculate the modified flow velocity $u^{\mu}$ using the Landau frame definition
\begin{align}
    T^{\mu\nu}u_\nu = \epsilon u^{\mu} \,.
\end{align}
$T^{03}/P_{eq}$  contains the directionality of the force as its sign changes with the force direction. Therefore we can use $T^{03}/P_{eq}$ and $\delta u/u << 1$ as a measure of self consistency. Here $P_{eq}$ is the equilibrium isotropic pressure defined by
\begin{equation}
    P_{eq} = \int d{\Xi} \frac{{\bm{p}^2}}{3} \, f_{eq} \,.
\end{equation}

 For the rest of our calculation, we use the conformal relation $\tau_R(\tau) = 5c/T(\tau)$, where $c \propto \frac{\eta_0}{s_0}$ is a dimensionless number which sets the initial viscosity of the system. In this study, we set the initial time to $\tau_0 = 0.1$fm and the initial temperature to $T_{0} = 500$ MeV. For the non-conformal computations, we set the mass to $m = 200$ MeV.

 The  Eq.\eqref{eq:E-Tau} can be solved using iterative techniques (A faster algorithm compared to the iterative technique used in previous studies is given in App.\eqref{sec:appIter}) and using the energy density-temperature relation,
\begin{equation}
   \mathcal{E}(T) = \frac{3T^4}{\pi^2} \lrb{ \frac{z^2}{2}K_2(z) + \frac{z^3}{6}K_1(z) }\,,
\end{equation}
where $z = m/T$ and $K_1,K_2$ are modified Bessel functions of the second kind. The relations for pressure Eq.\eqref{eq:PL-Tau},Eq.\eqref{eq:PT-Tau} can be evaluated once the temperature is obtained using the above procedure.

\subsection{Parametrised Force}
\subsubsection{Force Type 1}
\begin{figure}[!h]\label{Fig:Force}
  \includegraphics[width=.45\textwidth]{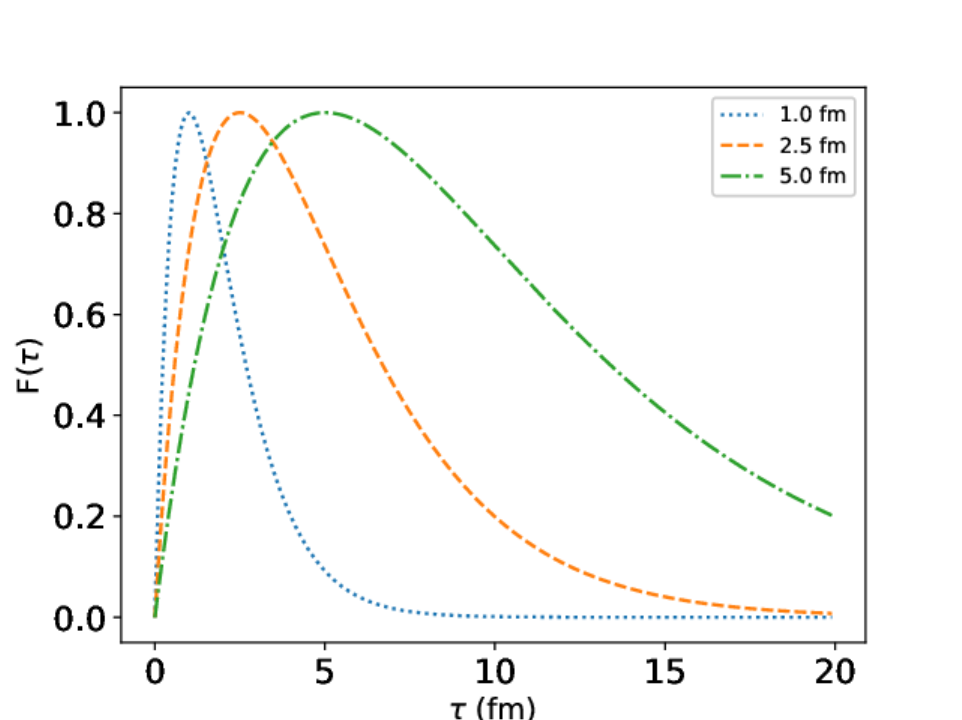}
\caption{Normalised force (max($F$) = 1) for different timescales. $\tau_F$ = 1 fm/c (dot), $\tau_F$ = 2.5 fm/c (dashed), $\tau_F$ = 5 fm/c (dot-dashed)}
\label{fig:Force}
\end{figure}

Motivated by the exponentially decaying fields in high-energy heavy-ion collisions we use the following parametrisation of the force
\begin{align}
    F(\tau) = \frac{\tau}{\tau_F}\exp{1-\frac{\tau}{\tau_F}},
\end{align}
where $\tau_F$ is a decay time scale.  The parameter $\tau_F$ is the proper time at which $\mathcal{F}$ takes its maximum value $\frac{\alpha}{ \tau_F}$. Here $\alpha$ has the dimension of momentum, $\frac{\alpha}{\tau_F}$ has dimensions of $p^2$(Force). By varying the value of $\alpha$ and $\tau_F$ we can control the strength and the duration of the force. A plot of the normalized force for various values of decay time is given in Fig.\eqref{fig:Force}.

\subsubsection{Force Type 2}
For exploratory purposes, we also try a constant force
\begin{align}
    F(\tau) = 1 ,
\end{align}
where $\mathcal{F}$ takes the maximum value $\mathcal{F}_{0}$. For ease of comparison, we rewrite $\mathcal{F}_{0} = \frac{\alpha}{ \tau_F}$. 

\section{Effect of force on fluid variables} {\label{sec:results}}

\subsection{Force Type  1}
\begin{figure}[p]
  \begin{minipage}{0.5\textwidth}
    \centering
    \includegraphics[width=0.72\textwidth]{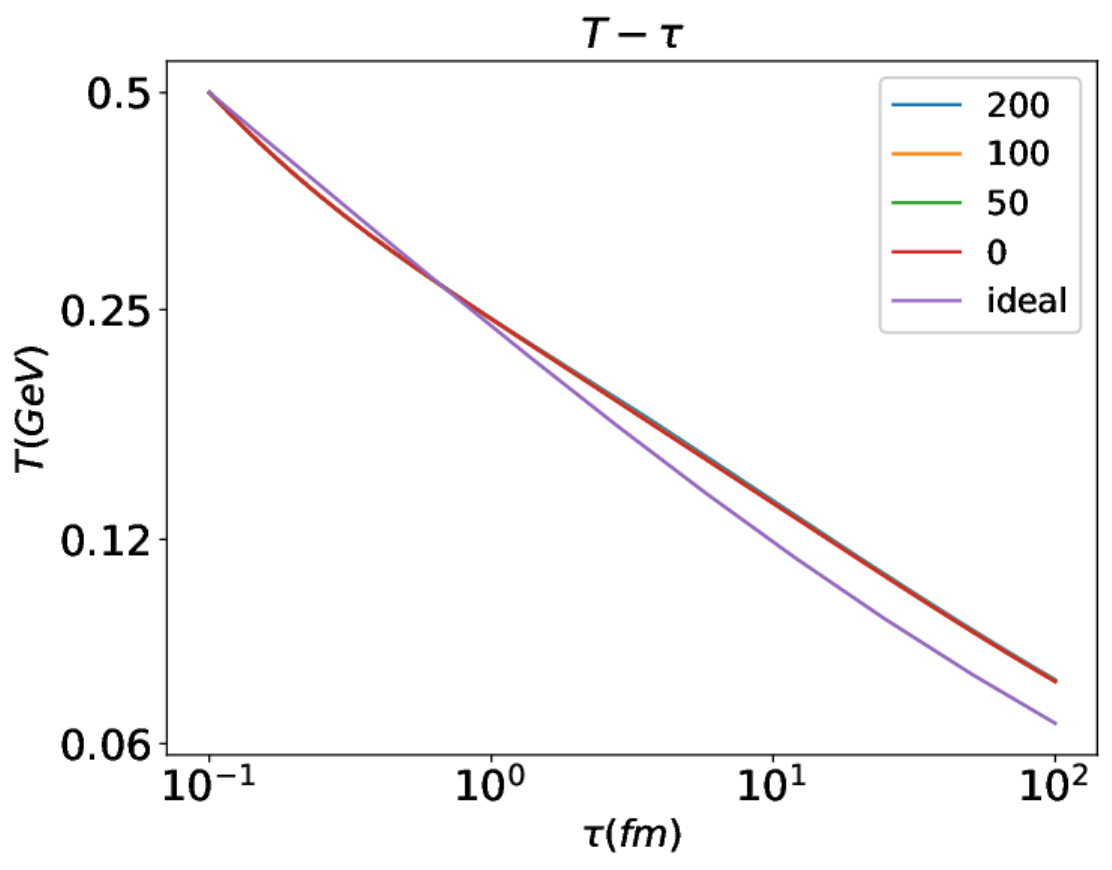}
     \includegraphics[width=0.72\textwidth]{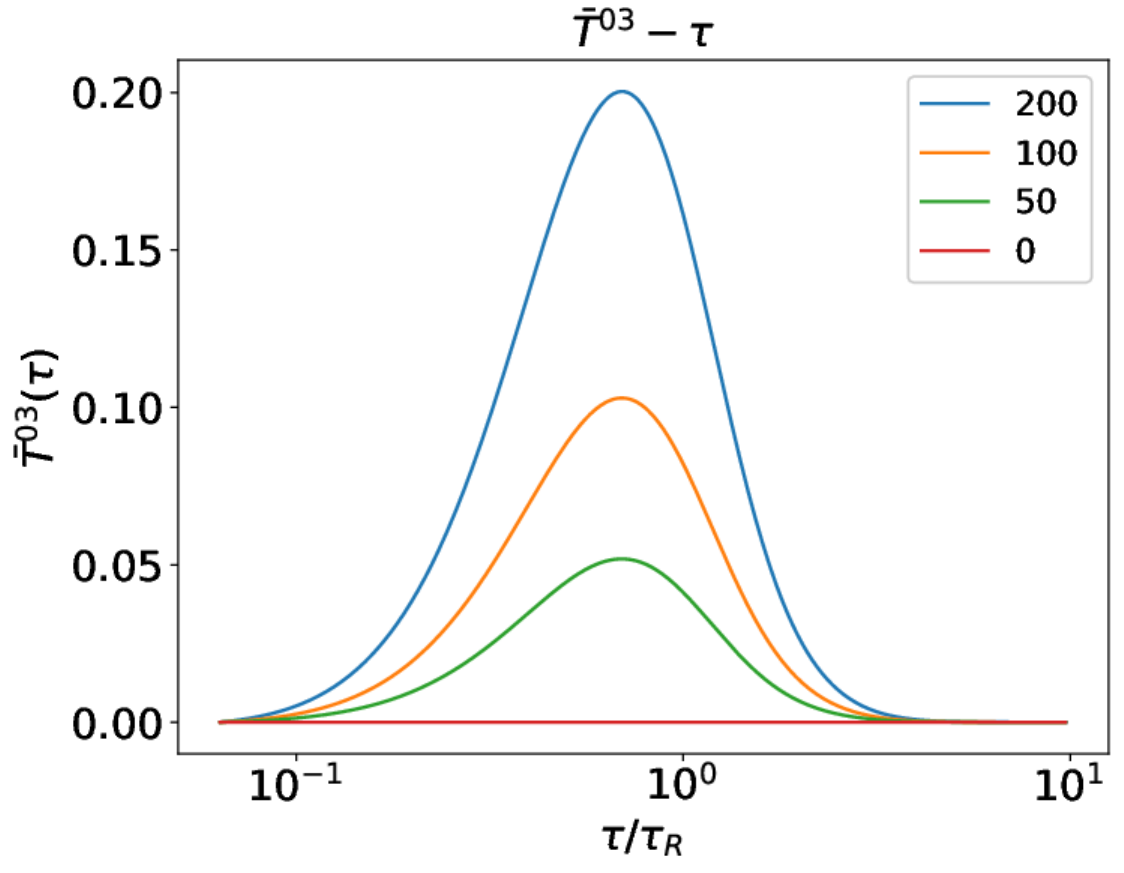}    
  \end{minipage}%
  \begin{minipage}{0.5\textwidth}
    \centering
    \includegraphics[width=0.72\textwidth]{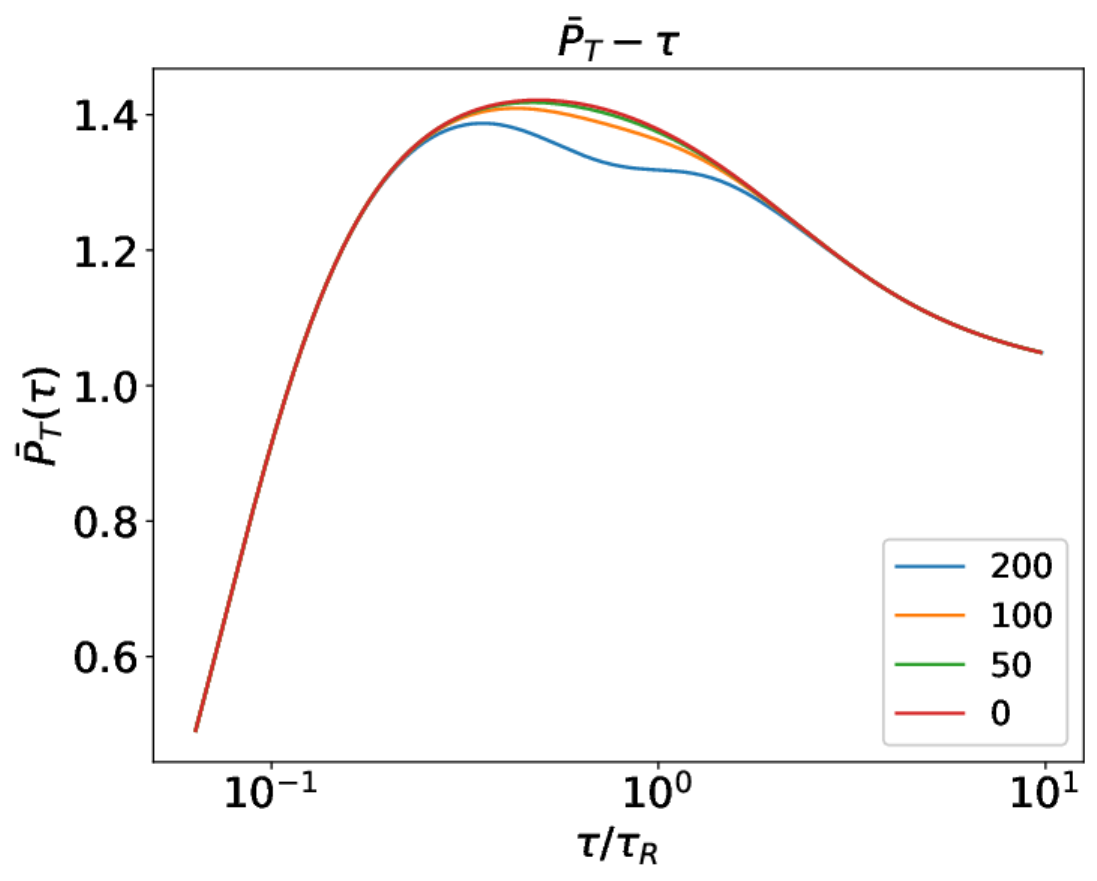}
   \includegraphics[width=0.72\textwidth]{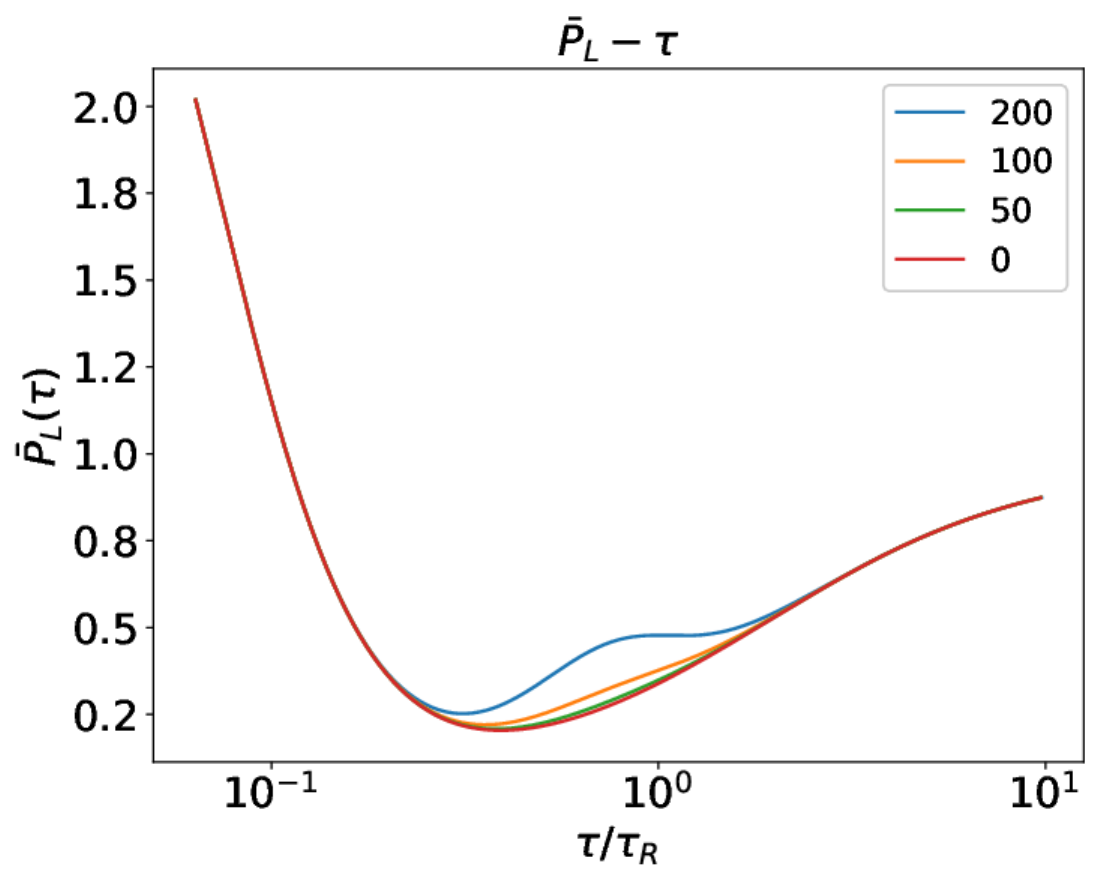}
  \end{minipage}
\begin{minipage}{\textwidth}
  \caption{Evolution of $T$, $\Bar{P}_L$, $\Bar{P}_T$, and $\Bar{T}^{03}$ for $\alpha = [200,100,50,0]$, $\tau_m = 1$, and for $\Bar{\Pi}_0 = 0$, $\Bar{\pi}_0 = -1$.}
  \label{fig:Exp_PL_1}
\end{minipage}
\end{figure}

\begin{figure}[p]
     \begin{minipage}{0.5\textwidth}
    \centering
    \includegraphics[width=0.72\textwidth]{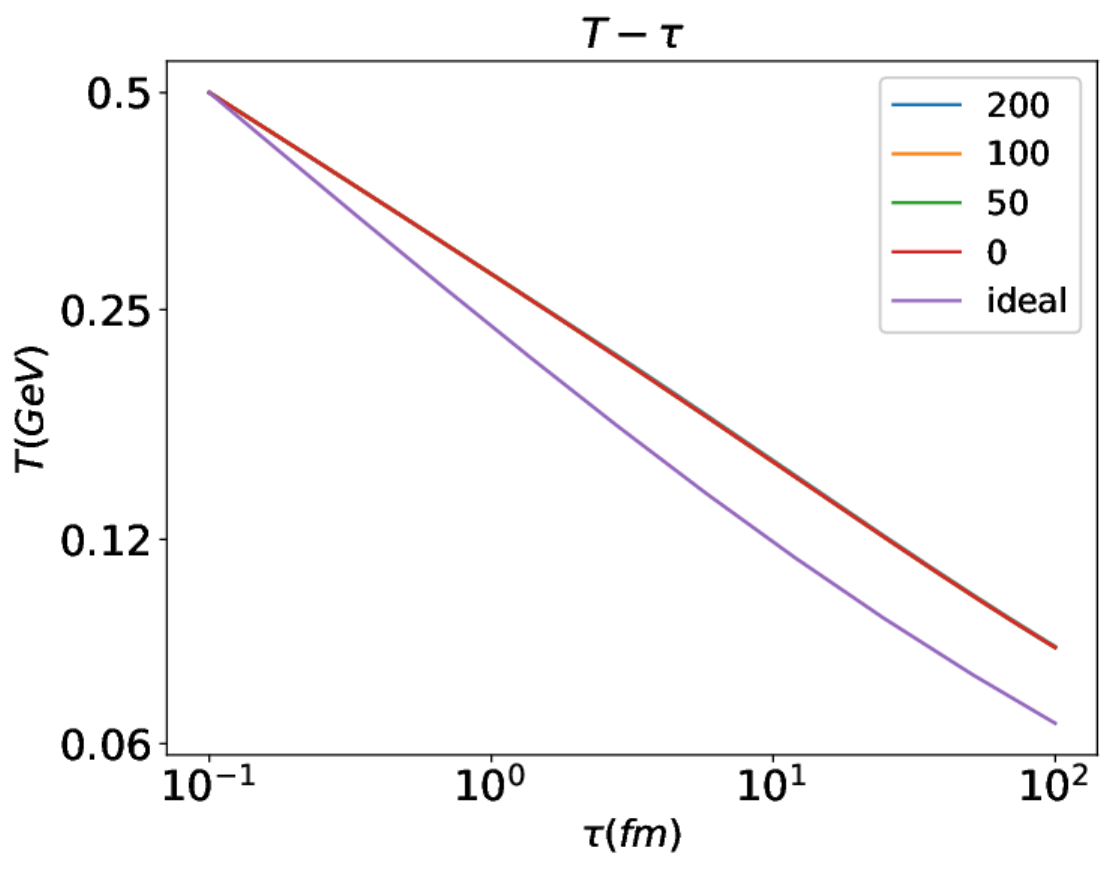}

     \includegraphics[width=0.72\textwidth]{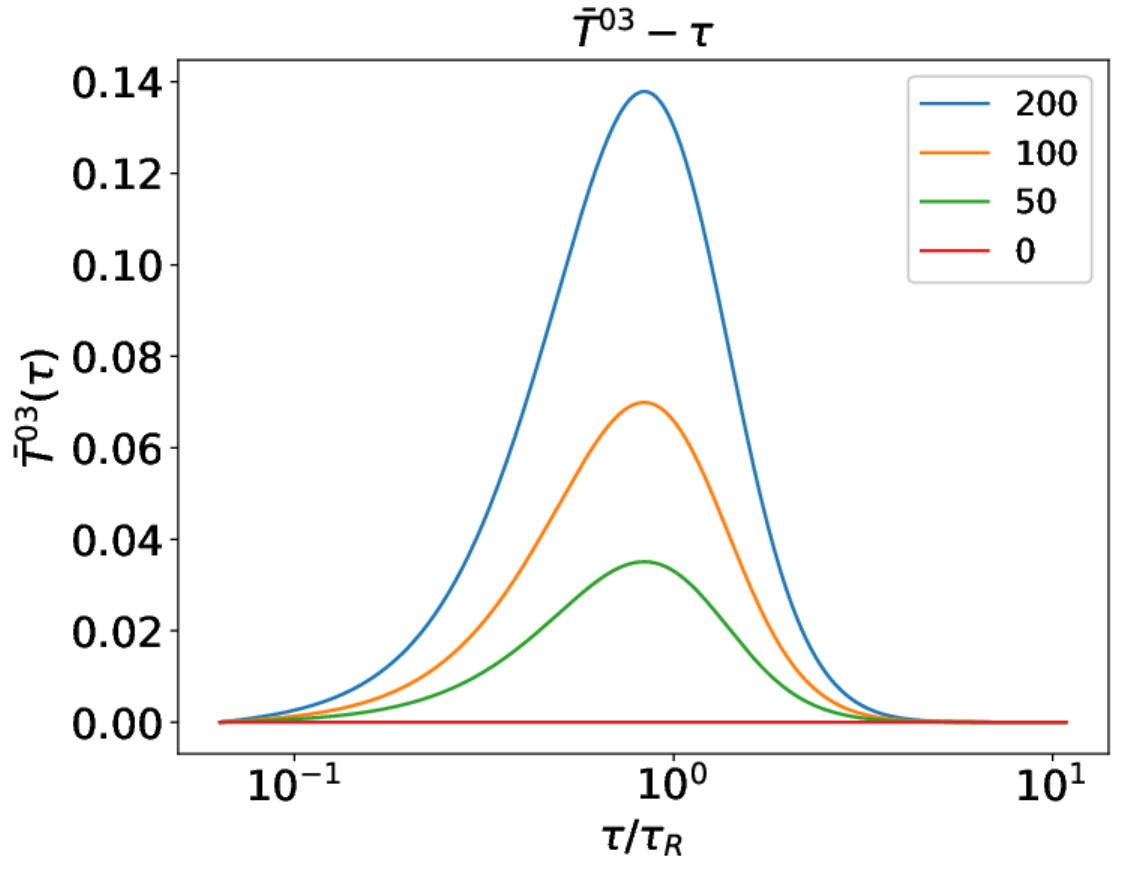}
 
  \end{minipage}%
  \begin{minipage}{0.5\textwidth}
    \centering   
  \includegraphics[width=0.72\textwidth]{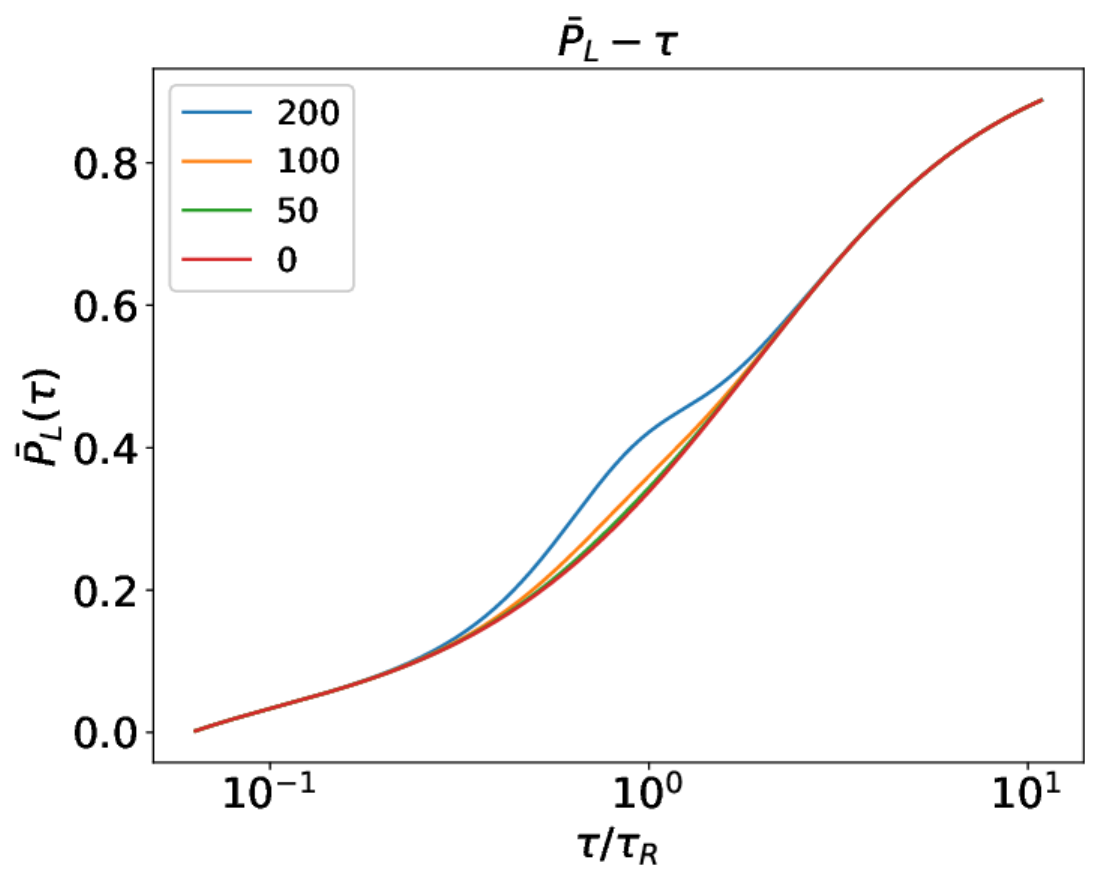}
   \includegraphics[width=0.72\textwidth]{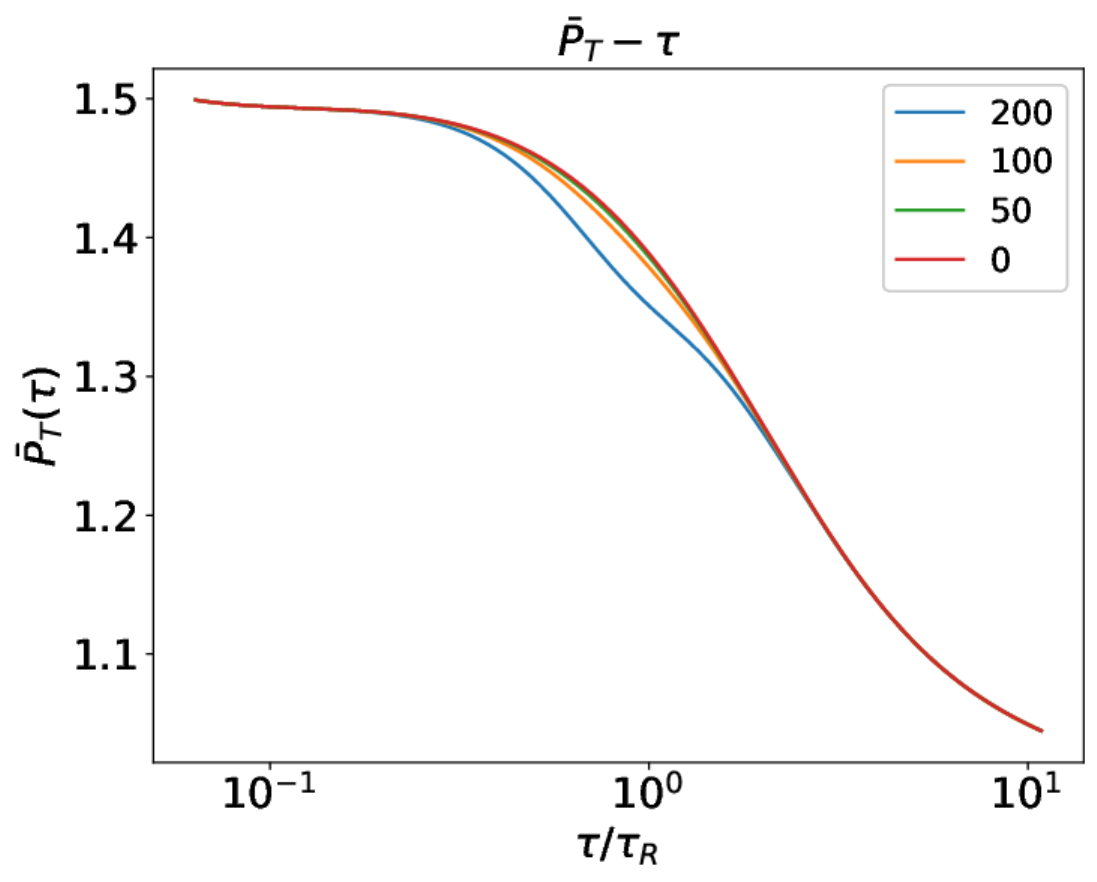}
  \end{minipage}
  \begin{minipage}{\textwidth}
  \caption{Evolution of $T$, $\Bar{P}_L$, $\Bar{P}_T$, and $\Bar{T}^{03}$ for $\alpha = [200,100,50,0]$, $\tau_m = 1$, and for $\Bar{\Pi}_0 = 0$, $\Bar{\pi}_0 = 0.99$.}
  \label{fig:Exp_PL_2}
  \end{minipage}
\end{figure}

\begin{figure}[p]
  \begin{minipage}{0.5\textwidth}
    \centering
    \includegraphics[width=0.72\textwidth]{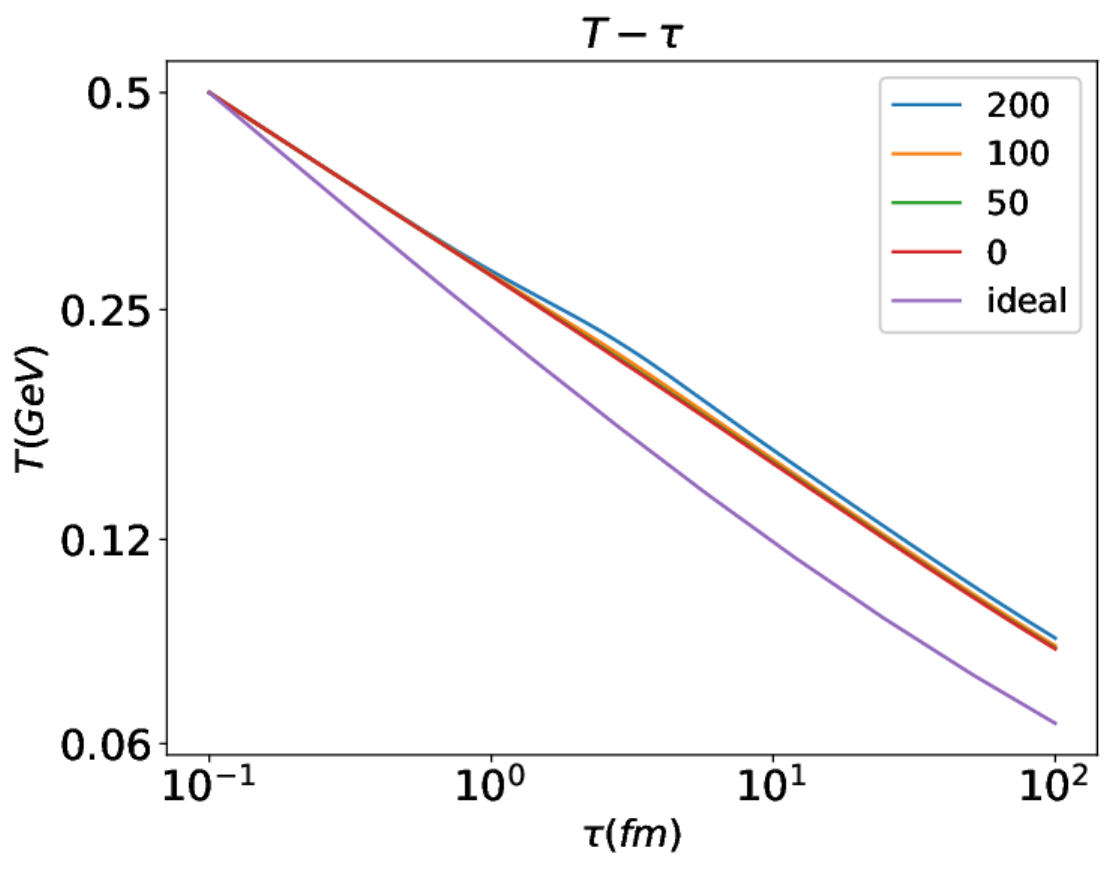}
   \includegraphics[width=0.72\textwidth]{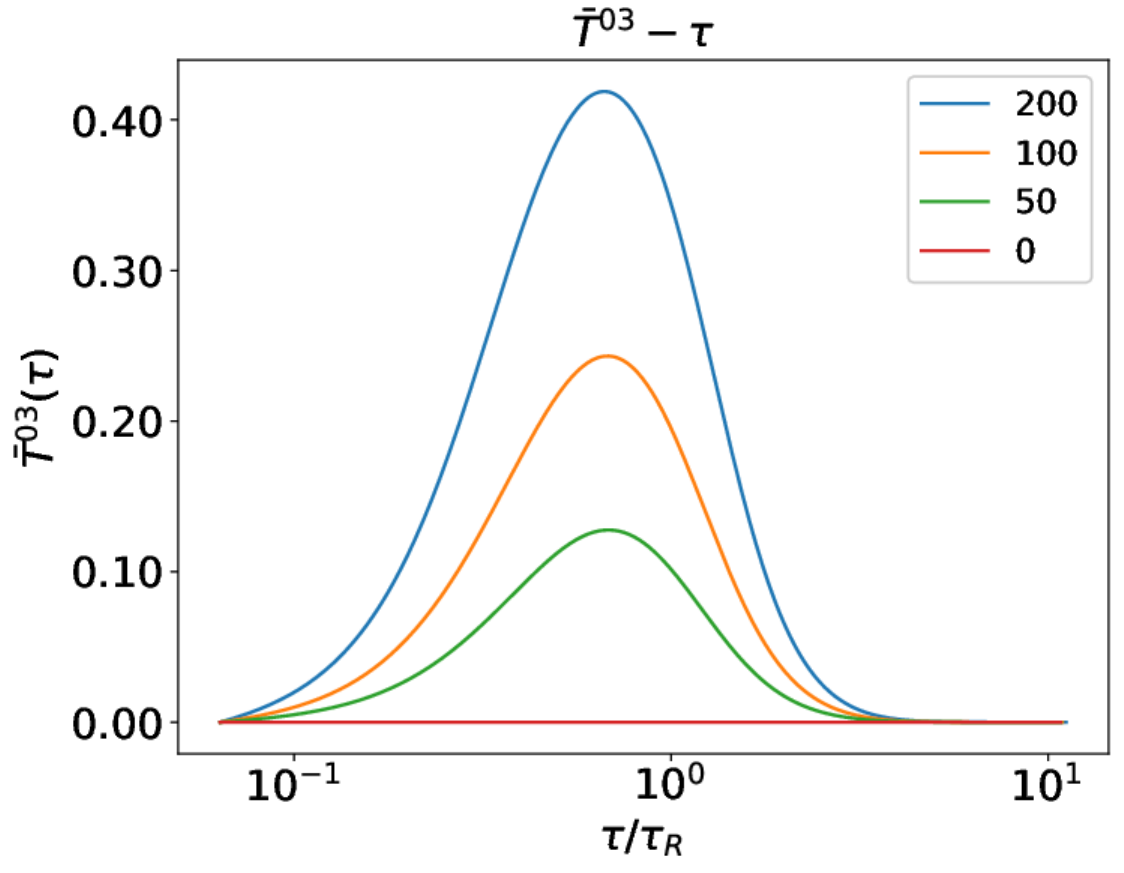}
  \end{minipage}%
  \begin{minipage}{0.5\textwidth}
    \centering   
     \includegraphics[width=0.72\textwidth]{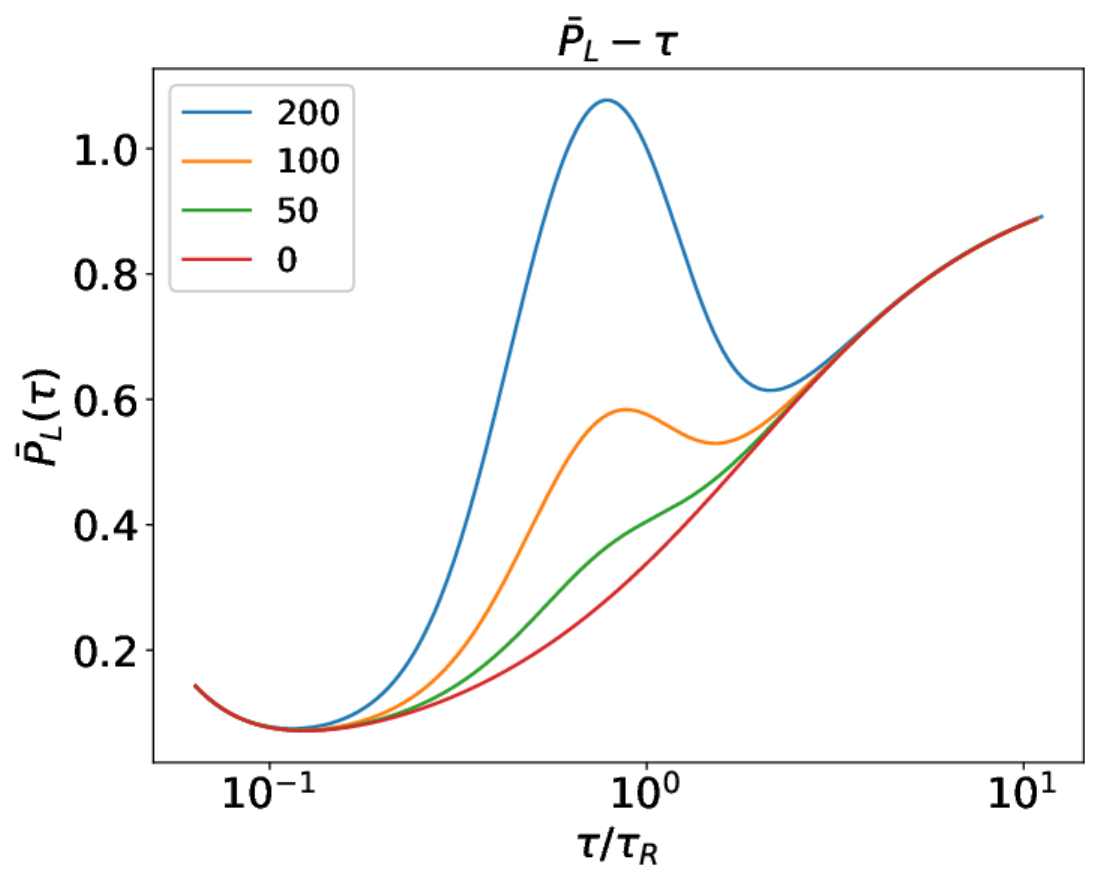}    
    \includegraphics[width=0.72\textwidth]{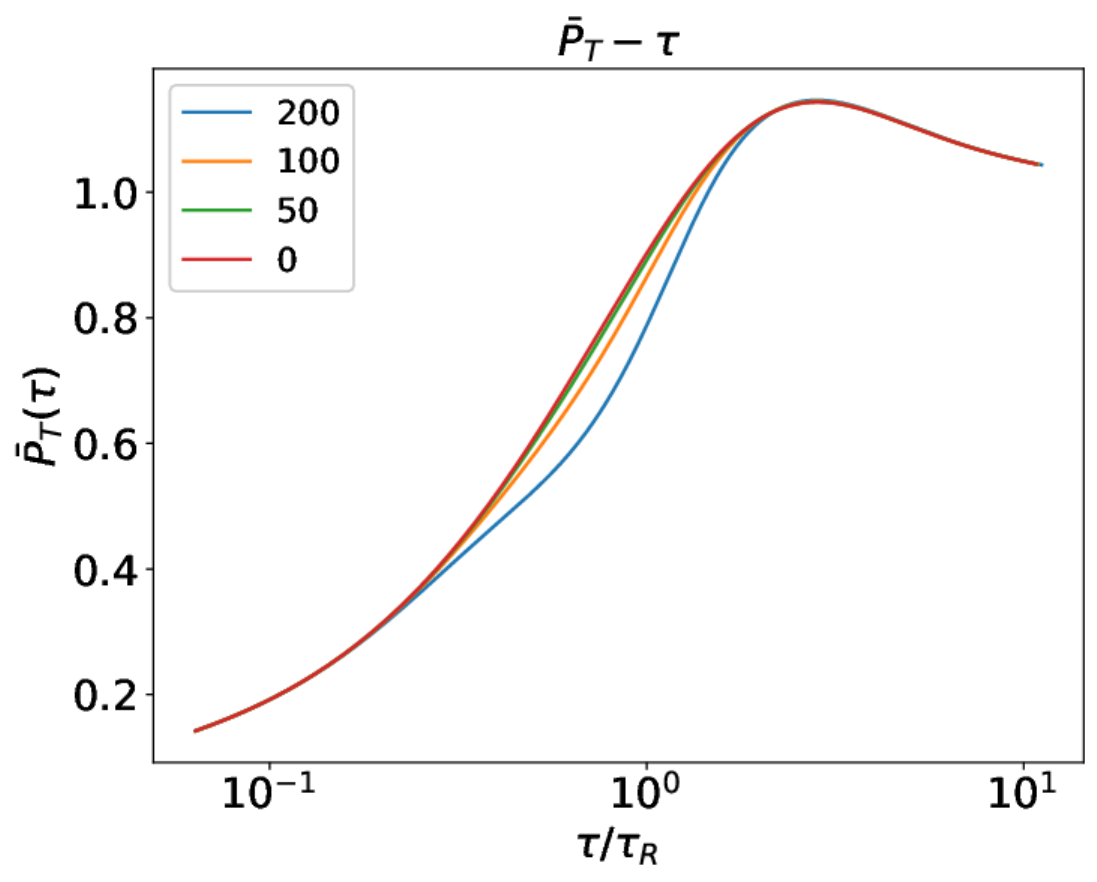}
  \end{minipage}
  \begin{minipage}{\textwidth}
    \captionof{figure}{Evolution of $T$, $\Bar{P}_L$, $\Bar{P}_T$, and $\Bar{T}^{03}$ for $\alpha = [200,100,50,0]$, $\tau_m = 1$ and for $\Bar{\Pi}_0 = 0.85, \Bar{\pi}_0 = 0$.}
    \label{fig:Exp_PL_3}
  \end{minipage}
\end{figure}

\begin{figure}[p]
  \begin{minipage}{0.5\textwidth}
    \centering
    \includegraphics[width=0.72\textwidth]{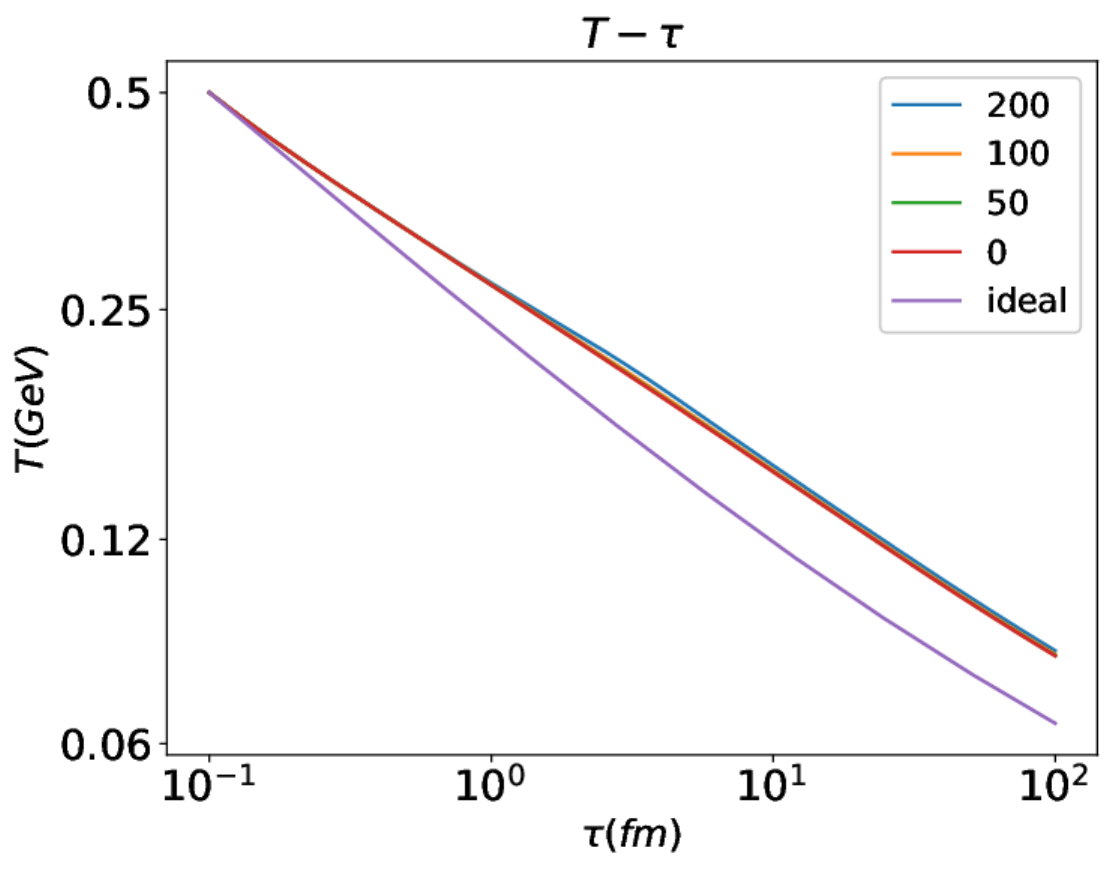}
     \includegraphics[width=0.72\textwidth]{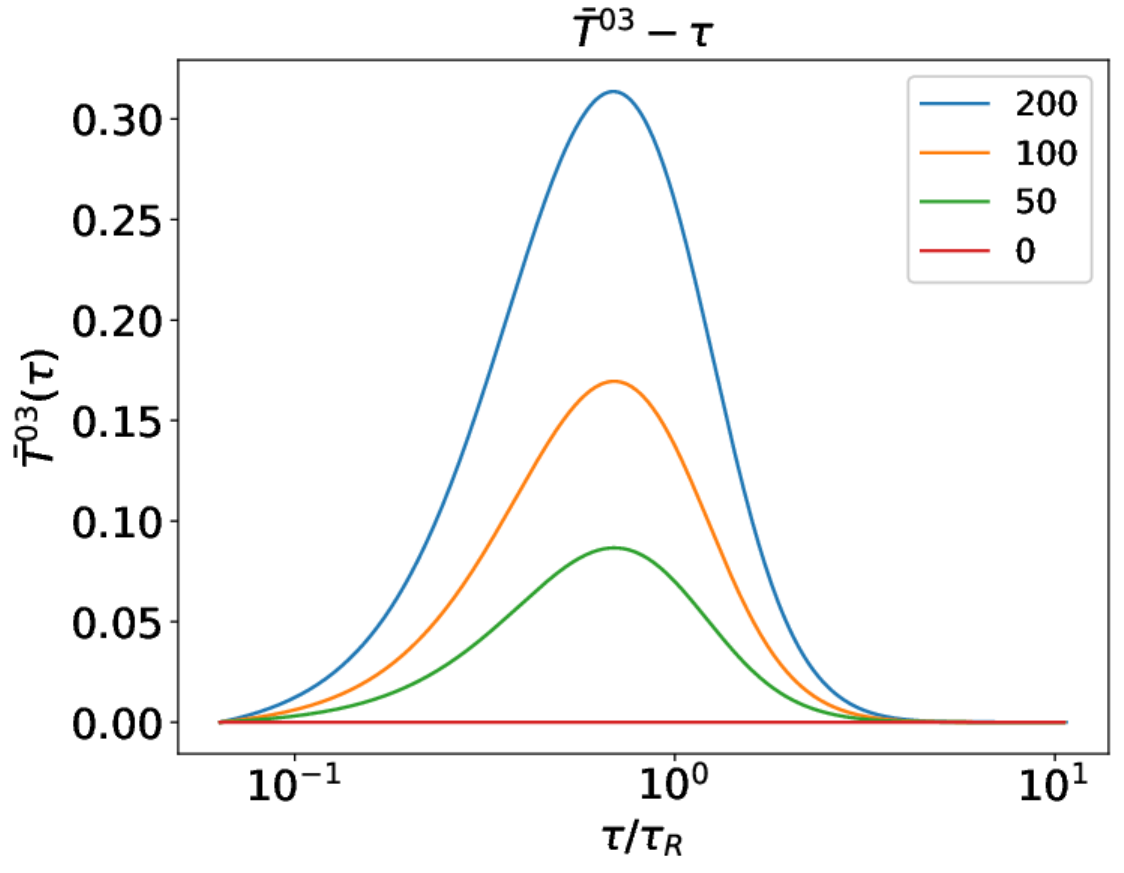}
  \end{minipage}%
  \begin{minipage}{0.5\textwidth}
    \centering    
    \includegraphics[width=0.72\textwidth]{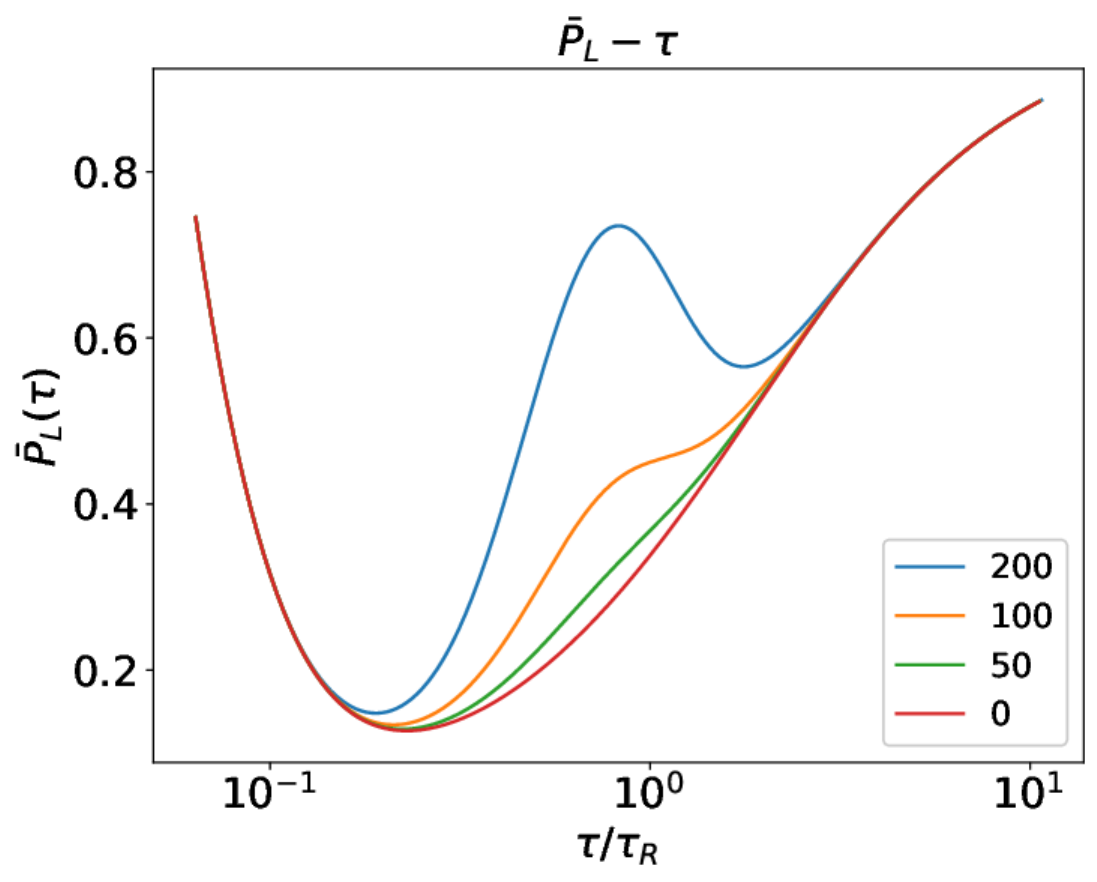}   
    \includegraphics[width=0.72\textwidth]{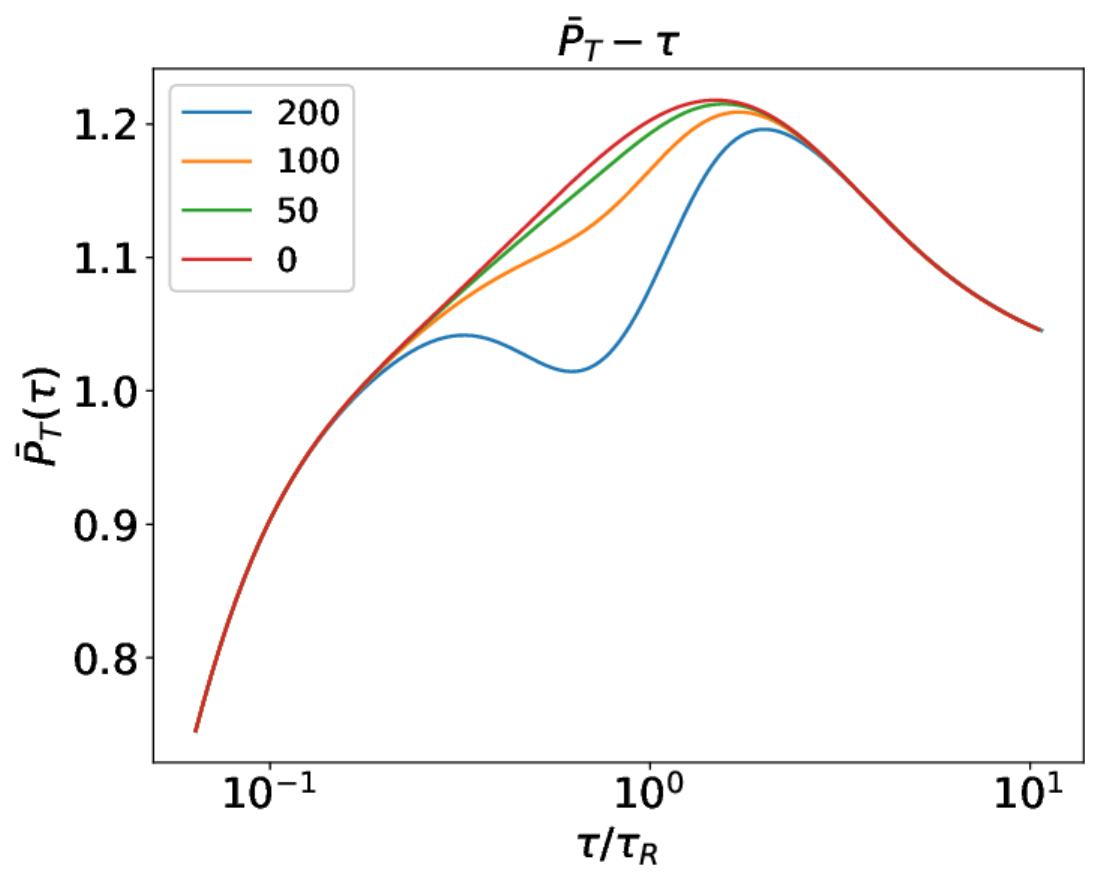}
  \end{minipage}
  \begin{minipage}{\textwidth}
    \captionof{figure}{Evolution of $T$, $\Bar{P}_L$, $\Bar{P}_T$, and $\Bar{T}^{03}$ for $\alpha = [200,100,50,0]$, $\tau_m = 1$ and for $\Bar{\Pi}_0 = 0.25, \Bar{\pi}_0 = 0$.}
    \label{fig:Exp_PL_4}
  \end{minipage}
\end{figure}

 The application of an external force should change the particle momenta and therefore could have observable effects on pressure. We explore these effects in Fig.[\ref{fig:Exp_PL_1}-\ref{fig:Exp_PL_4}] for various values of $m/\Lambda_0$. The figures [\ref{fig:Exp_PL_1}-\ref{fig:Exp_PL_4}] show the proper time evolution of temperature  (top left panel), transverse pressure (top right panel), longitudinal pressure (bottom right panel) and the scaled off-diagonal component $T^{03}/P_{eq}$ (bottom left panel) of the energy-momentum tensor for various values of initial anisotropy. Different lines in these figures correspond to various values of $\alpha$. Fig [\ref{fig:Exp_PL_1}] shows the time evolution of temperature for negative $\xi_0$ values whereas Fig[\ref{fig:Exp_PL_2}] is for a sharply peaked $p_z$ distribution. Fig.\eqref{fig:Exp_PL_3},\eqref{fig:Exp_PL_4} is for isotropic initial distribution but for varying $m/\Lambda_0$ values.  In each figure showing the temperature evolution, the red line corresponds to ideal hydro evolution,
and green, orange, and blue correspond to Boltzmann solution with  $\alpha= 200,\,100,\,50,\,0$ respectively. The same color code is used in $\Bar{P}_L,\Bar{P}_T$ , and $\Bar{T}^{\tau z}-\tau$ evolution.

%\clearpage

\clearpage

\clearpage

It is seen that for the magnitude of the force considered in this study, there is a marginal impact on the evolution of energy density (temperature).  A possible explanation is the following: the expansion cools down the system whereas force, in the presence of dissipation, should increase the temperature of the system. The effect is small because we are assuming the force and the correction to the fluid velocity is small. 

 It is also seen that $\Bar{P}_L$ increases while $\Bar{P}_T$ correspondingly decreases during the time period in which the force acts. $P_L$ increases as it is driven by an external force, which in turn drives the transverse pressure down due to the constraint $\mathcal{E} - 2P_T -P_L = \langle m^2 \rangle$. We see that the long-term behavior of the evolution is such that the system relaxes back to the zero-force curve, and loses memory of the variations due to force. 

From Fig.\eqref{fig:Exp_PL_1} and Fig.\eqref{fig:Exp_PL_2} we see that the effect of longitudinal force is diminished when there is a large initial shear anisotropy. That means that the predominant factor driving the evolution of the system is the shear force. On the other hand in Fig.\eqref{fig:Exp_PL_3} and Fig.\eqref{fig:Exp_PL_4} when the shear anisotropy is low, we see that the effect of force is pronounced and the effect of bulk anisotropy is subdominant. 

\subsection{Force Type 2}
\begin{figure}[!ht]
  \includegraphics[width=0.37\textwidth]{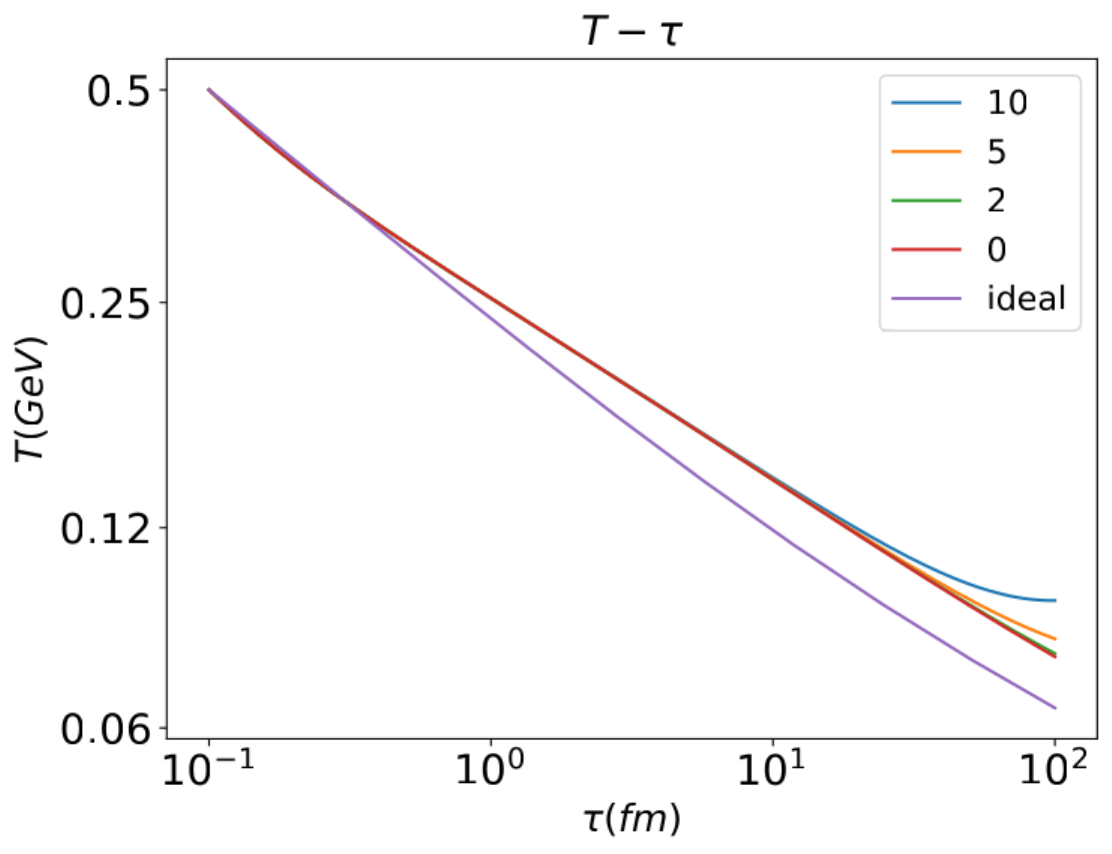}\\
  \includegraphics[width=0.37\textwidth]{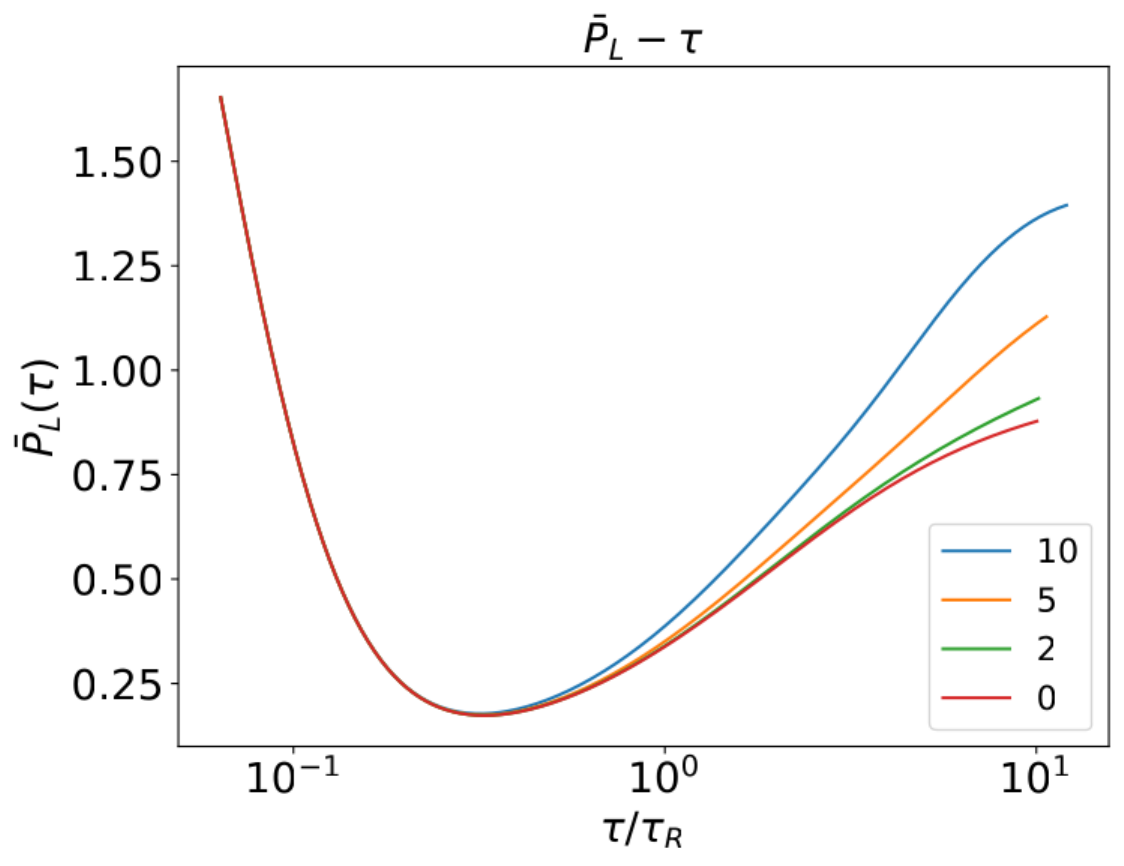}\\
  \includegraphics[width=0.37\textwidth]{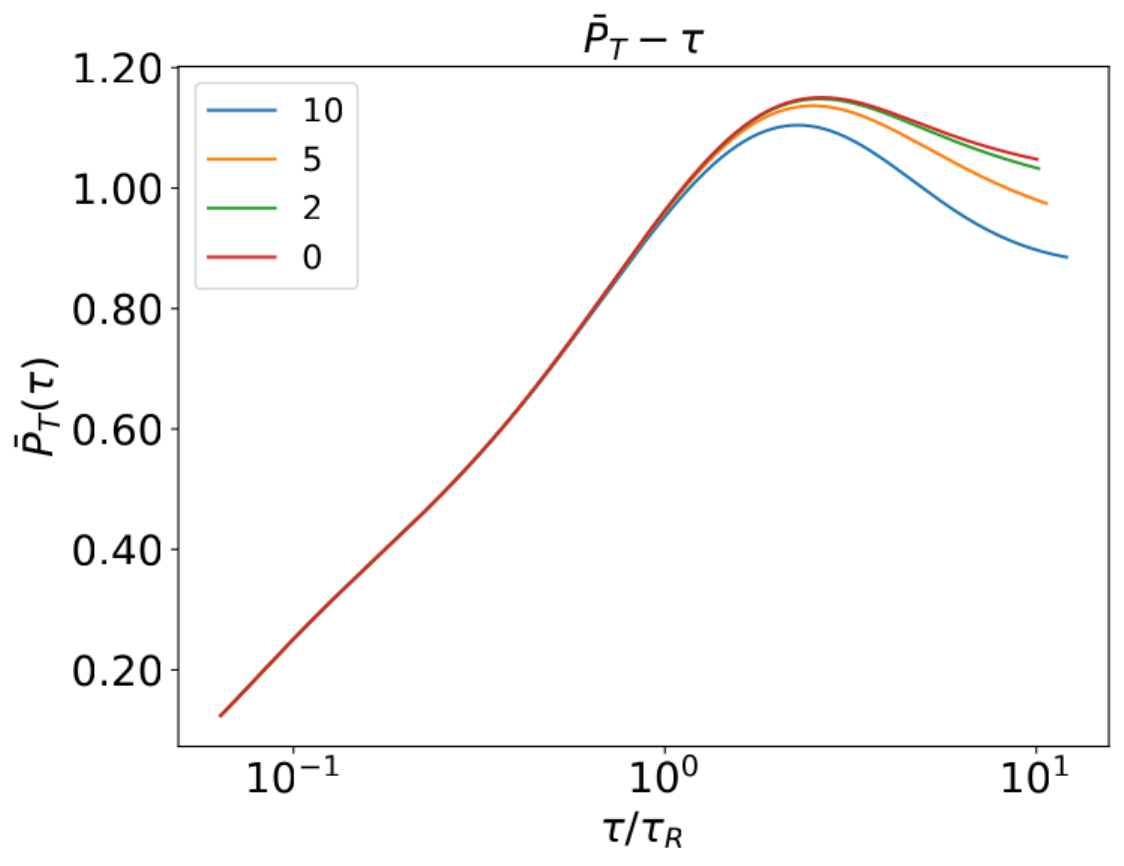}\\
  \includegraphics[width=0.37\textwidth]{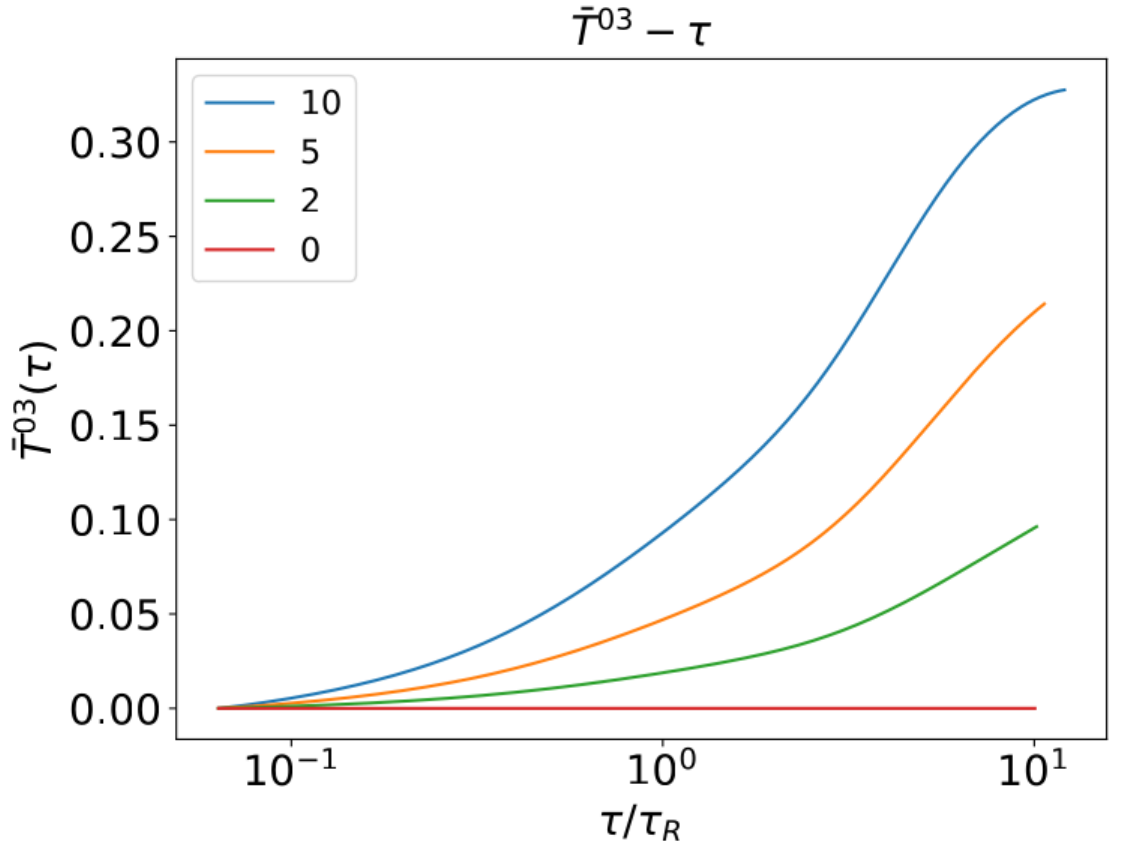}\\
\caption{Evolution of $T$,$\Bar{P}_L$,$\Bar{P}_T$ and $\Bar{T}^{03}$ for $\alpha = [10,5,2,0]$,$\tau_m = 1$ and for $\Bar{\Pi}_0 = -0.37,\Bar{\pi}_0 = -1$}
\label{fig:Con_PL_2}
\end{figure}

In the case of constant force, the force-to-temperature ratio cannot be kept constant throughout the evolution of the system. So, we use smaller $\alpha = 2, 5, 10 $ (MeV) values to make the corrections small. We see again that the variation in temperature is marginal except for late times when the force is comparatively larger than the effective temperature. The variations in the bulk observables are also only noticeable at late times. The qualitative behavior of the bulk observables remains the same as that for the case of the decaying force. We show the results for a constant force for $\Bar{\Pi}_0 = -0.37$,$\Bar{\pi}_0 = -1$ in Fig.\eqref{fig:Con_PL_2}.

\subsection{Attractor Behaviour: Non-conformal}

\begin{figure}[h]
    \centering
    \includegraphics[width=.40\textwidth]{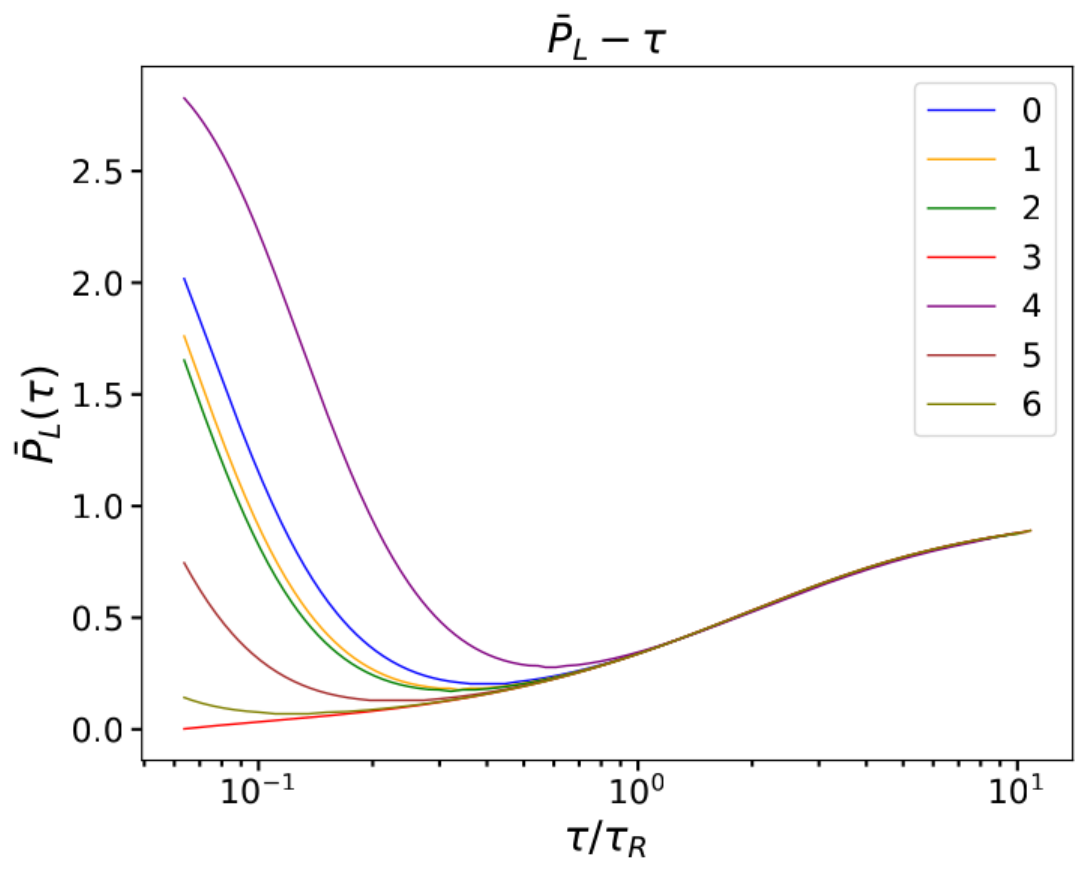}\\
  \includegraphics[width=.40\textwidth]{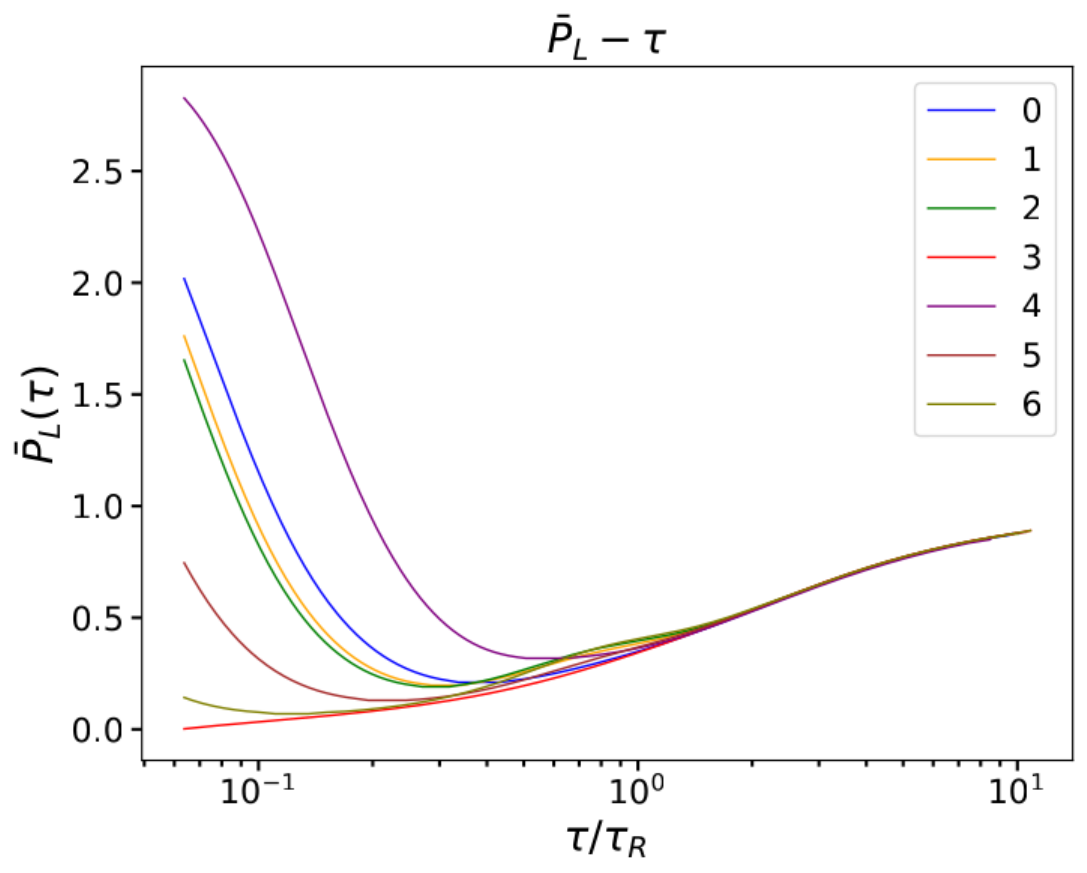}\\
  \includegraphics[width=.40\textwidth]{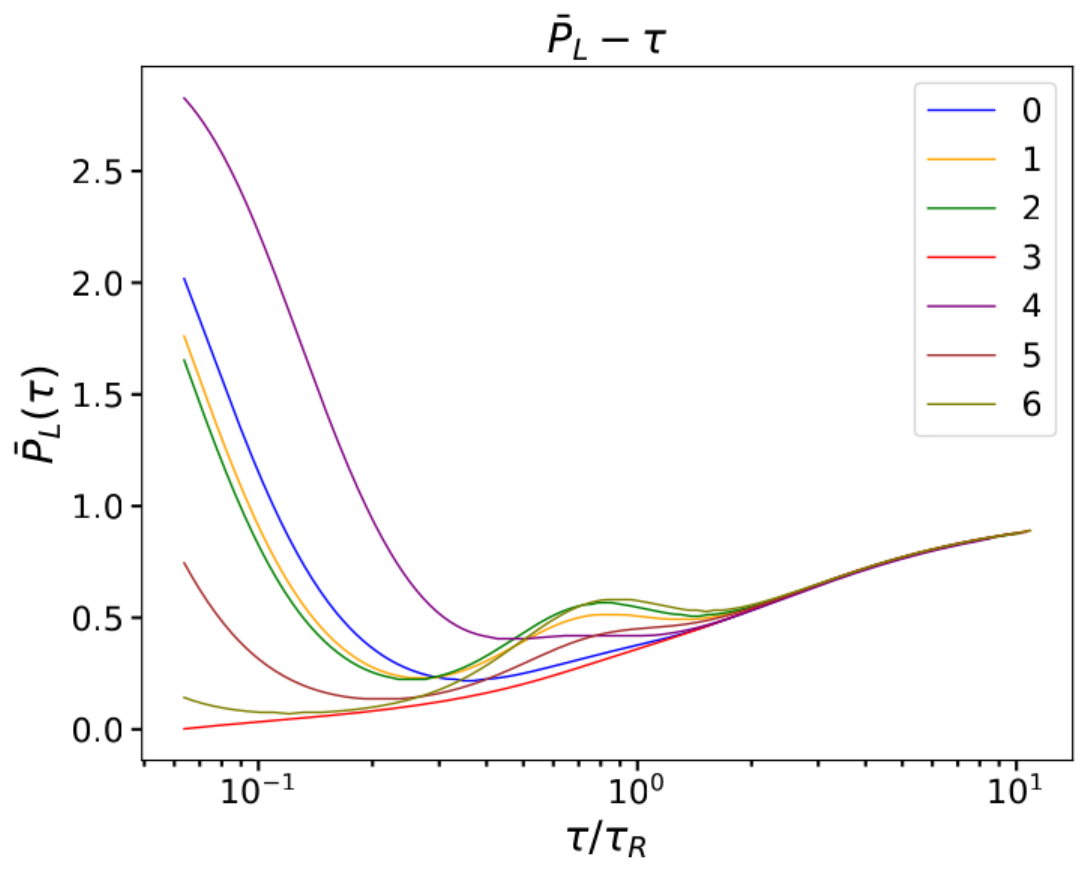}
    \caption{Evolution of $\Bar{P}_L$ for $\alpha = [0,50,100]$,$\tau_m = 1$ and for various initial conditions}
    \label{fig:ConAttr}
\end{figure}

It was reported that for non-conformal systems a universal attractor was not obtained for fluid variables other than for scaled longitudinal pressure $\Bar{P}_L=P_L/P$ \cite{Jaiswal:2022mdk}. The authors also computed the bounds for $\Pi/P$ (scaled bulk pressure) and $\pi/P$ (scaled sheer pressure) so that the positivity of the distribution function, the transverse and the longitudinal pressure is maintained. Following their findings we use the same set of initial parameters given in Table \eqref{tab:Init_Pi_Jaiswal}. The corresponding values for $m/\Lambda_0$, $N_0$ and $\xi_0$ are reported in Table.\eqref{tab:Init_Jaiswal}. With this, we check how the attractor behavior is modified by an external force. In Fig.\eqref{fig:ConAttr} the scaled longitudinal pressure $\Bar{P}_L$ is plotted for three different force $\alpha=0$ (top panel), $\alpha=50$ (middle panel), and $\alpha=100$ (bottom panel). In each subplot, the different colored lines represent different anisotropies as given in Table.\eqref{tab:Init_Jaiswal}. For zero force we reproduce the results in the the previous study \cite{Jaiswal:2022mdk}. However, for finite forces, only a late-time attractor is present.

\subsection{Attractor Behaviour: Massless Limit}

\begin{figure}[!ht]
\includegraphics[width=.40\textwidth]{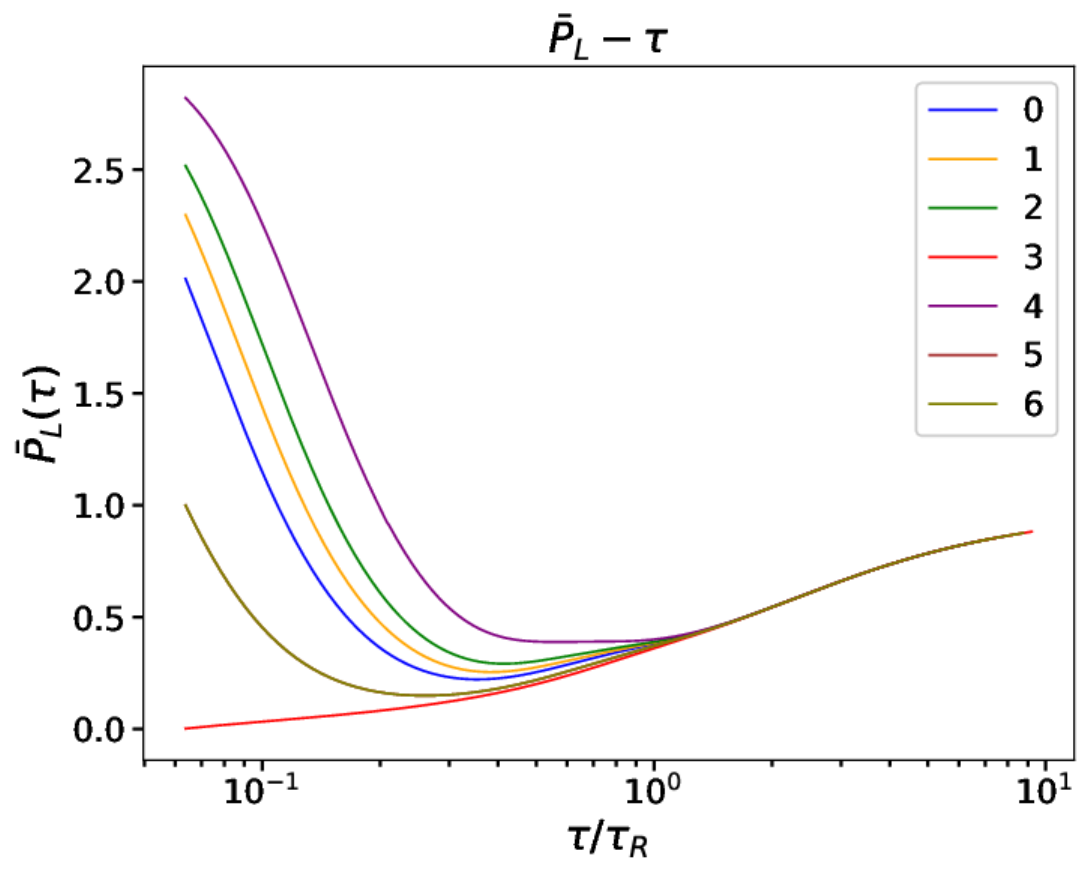}\\
\includegraphics[width=.40\textwidth]{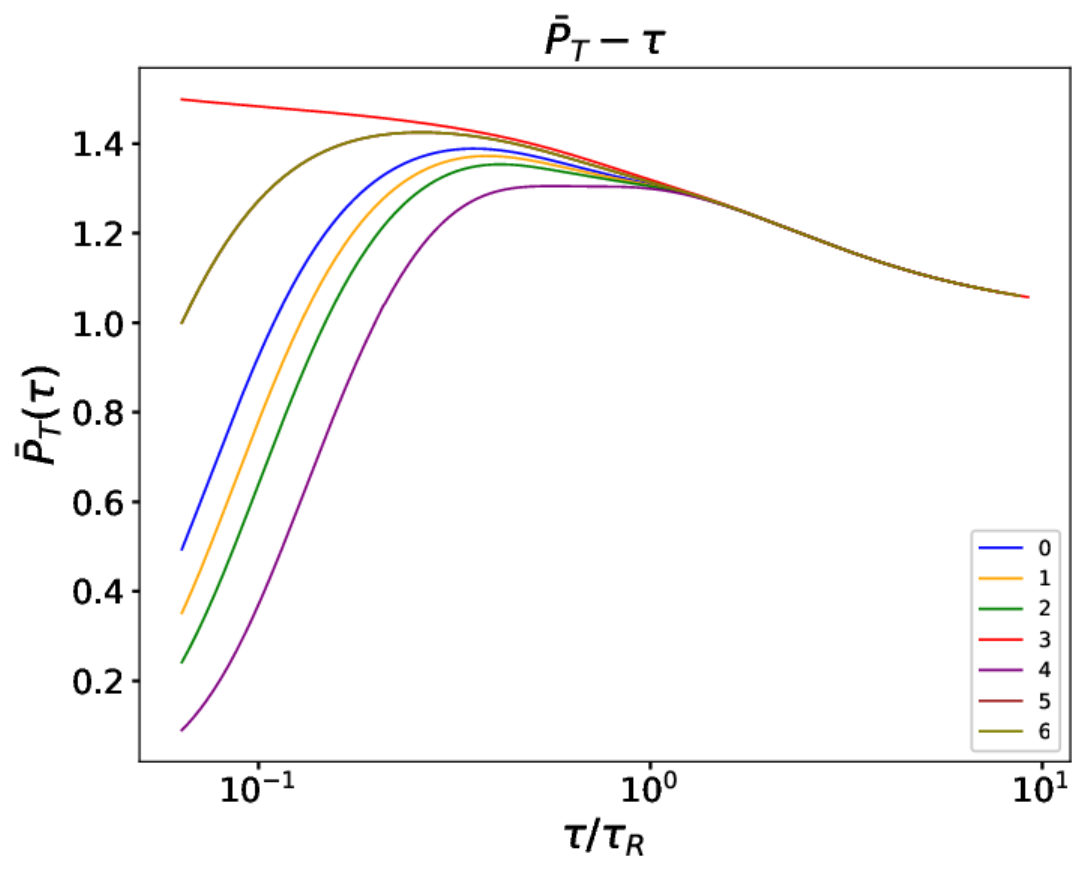}
\caption{ Scaled Transverse and longitudinal pressure plotted for varying initial conditions (given in Table.\eqref{tab:Init_Conf}) in the presence of an external force $\alpha = 100$ with m=0.}\label{fig:Conf_P}
\end{figure}

Systems with conformal symmetry were shown to exhibit universal early time attractor behavior for a variety of system configurations and flow profiles \cite{Heller:2015dha,Mitra:2022uhv,Florkowski:2017jnz,Denicol:2019lio,Romatschke:2017acs,Dash:2022agb}. Here we check if early-time attractors can be reproduced for systems under external forces. In Fig.\eqref{fig:Conf_P} the transverse and longitudinal pressure is plotted for varying initial conditions (given in Table.\eqref{tab:Init_Conf}) in the presence of an external force in the massless limit. We observe that even when the system is subjected to an external force it shows an attractor in $\bar{P}_{L}$ and $\bar{P}_{T}$ much earlier than the massive case (as compared to Fig.\eqref{fig:ConAttr}). We note however that this is subject to the condition that the force has died out much before equilibration is achieved.

The equilibration rate is mostly determined by the exponentially decaying damping factor $D(\tau,\tau')$  given in Eq.\eqref{Eq:Damp}. One can see from Eq.\eqref{eq:E-Tau}, Eq.\eqref{eq:PL-Tau} and Eq.\eqref{eq:PT-Tau} that the decay rate is determined by the interplay between the damping factor and the $H_{\mathcal{E}/P_L/P_T}$ functions. It is observed from the $\tau-T$ graphs that the force has negligible impact on the temperature evolution and, therefore, doesn't modify the damping function(as $\tau_R \sim 1/T$). However, it significantly modifies $H_{P_L/P_T}$ functions. Therefore, the relative change in behavior is due to the higher sensitivity of these functions on mass and force.  Within the approximation scheme employed, in the massless limit, the system regains conformal symmetry. A combination of these factors gives rise to the observed behaviour.

\section{Summary and Conclusion}{\label{sec:summary}}
 We solve the RTA Boltzmann equation incorporating the external force and study the dynamics of the system far away from local equilibrium under Bjorken flow. We also study the attractor behavior of a system undergoing Bjorken expansion in the presence of external longitudinal force. For this exploratory study, we consider a constant force and time-parameterized force which initially peaks and then exponentially decays afterward. We treat this force as a small perturbation so that the overall flow profile remains unchanged. For a constant force the temperature evolution was observed to vary at late times while for the time-varying force considered here, no significant change was observed for temperature evolution. The longitudinal pressure was observed to increase as it is driven by the external longitudinal force which in turn drove the transverse pressure down due to the constraint $\mathcal{E} - 2P_T -P_L = \langle m^2 \rangle$. The force was also observed to have a larger effect for smaller-scaled mass values.

We explored the attractor behavior by considering the time evolution of scaled longitudinal and transverse pressure for various initial anisotropies. For conformal systems an early time attractor was observed for various bulk scaled observables \cite{Heller:2015dha}. Whereas in the non-conformal case \cite{Jaiswal:2022mdk} an early time attractor was observed only for scaled longitudinal pressure. The addition of force however modifies this behaviour and an attractor is only found at late times.  In the conformal limit, we observed that, within our approximation scheme, the attractor for scaled longitudinal pressure reappears even in the presence of an external force. This behavior is due to the non-trivial interplay between the damping function $D(\tau,\tau')$ and $H_{P_L/P_T}$.
The results in this work were obtained by assuming that the system obeys approximate Bjorken symmetry. However, the non-diagonal terms $T^{03}$ become comparable to $\mathcal{E}$ and $\mathcal{P}$ for large values of force. Therefore, a proper quantitative estimate of the evolution of the bulk properties cannot be obtained using the techniques used in this paper.

For future work, one can explore the effect of force on non-Bjorken flows like Gubser and even arbitrary flow profiles. One can consider more realistic electromagnetic forces and a self-consistent solution can be obtained by considering a $1+1$ dimensional expansion. One can also solve for systems under a magnetic field which would necessarily include transverse dynamics and momentum dependant forces, 
Further, one can explore the effects of Chromo-Electric and Chromo-magnetic fields which would require modifications in the collision kernel.

\section{Acknowledgements}
AP acknowledges the CSIR-HRDG financial support. RG  and VR  acknowledge support from the DAE, Govt. of India. RG also acknowledges fruitful discussions with Sunil Jaiswal during the early stages of this work.

\appendix

\section{RTA Boltzmann Equation with Force}
\label{sec:appA}

Boltzmann equation would then read
  \begin{align}
       \mathcal{F}& \tau\pdv{f}{w} + \pdv{f}{\tau}  + \frac{\delta u_z p^z}{u_\tau p^\tau}\pdv{f}{z}  \\&= -\frac{p^\tau}{u_\tau p^\tau} \frac{\lrb{f-f_{eq}}}{\tau_R}  - \frac{\delta u \cdot p}{u_\tau p^\tau}\frac{\lrb{f-f_{eq}}}{\tau_R}  \, .
 \end{align}
 If we now assume that $\delta u/u_{\tau} <<1$, we can ignore the $\delta u$ terms from the above equation and get
  \begin{align}\label{Eq:BjBolF}
      \lrb{\mathcal{F}(\tau) \tau\pdv{}{w} + \pdv{}{\tau}}   f =  -\frac{\lrb{f-f_{eq}}}{\tau_R}  \, .
 \end{align}
 For the consistency of the above approximation, we need to keep the strength of the force sufficiently small so that the background flow is still close to Bjorken.
 \subsection{Change of Variables}\label{Ap:ChofVar}

 For computational convenience we write the force $\mathcal{F}$ as 
 \begin{equation}
     \mathcal{F} = \frac{\alpha}{\tau_F}F(\tau)\,.
 \end{equation}
 We then rewrite the Boltzmann equation as
   \begin{align}
      \lrb{\frac{\alpha}{\tau_F}F(\tau) \tau\pdv{}{w} + \pdv{}{\tau}}   f =  -\frac{\lrb{f-f_{eq}}}{\tau_R}  \, .
 \end{align}
 To convert Eq.\eqref{Eq:BjBolF} to a simpler form, we divide both sides by $F(\tau)\tau /\tau_F$ getting
   \begin{align}
      \lrb{\alpha \pdv{}{w} + \frac{\tau_F}{F(\tau) \tau}\pdv{}{\tau}}   f =  -\frac{\tau_F}{\tau_R}\frac{\lrb{f-f_{eq}}}{F(\tau) \tau }  \, .
 \end{align}
We define the variable $s$
\begin{align}
    s(\tau) &= \int_{0}^{\tau}F\lrb{\tau' }\frac{\tau'}{\tau_F}  d\tau',
\end{align}
and rewrite the proper time derivative to get Eq.\eqref{eq:BoltzForce},
  \begin{align}
      \lrb{\alpha \pdv{}{w} + \pdv{}{s}}   f =  -\frac{\tau_F}{\tau_R}\frac{\lrb{f-f_{eq}}}{F(\tau) \tau }  \, .
 \end{align}

 We now use the standard technique of characteristics for solving partial differential equations and define new variables $(w, s)\rightarrow (r, s)$
 \begin{align}
     r &= w - \alpha s, \\
     s &= s,
 \end{align}
 and obtain the equation
\begin{align}
      \pdv{ f(r,s,p_T)}{s}   =  -\frac{\tau_F}{\tau_R}\frac{\lrb{f-f_{eq}}}{F(\tau(s)) \tau(s) }  \, .
 \end{align}
%%%----------------------------------------------------------------------------------------------------------------------------

 \section{Integrals}\label{App:Int}
In this section we give the explicit form of the $H$ functions appearing in the equation for $\mathcal{E}$, $P$ etc given in Eq.\eqref{Eq:IntEPT}
 We define the integrals,
\begin{align}
    \Tilde{H}^{F}_{\epsilon}\lrrb{s',s,m,\alpha} = \int du\, u^3\, e^{-\sqrt{u^2 + z^2}}H^{F}_{\epsilon} (u,z,s,s',\alpha), \\
    \Tilde{H}^{F}_{L}\lrrb{s',s,m,\alpha} = \int du\, u^3\, e^{-\sqrt{u^2 + z^2}}H^{F}_{L} (u,z,s,s',\alpha), \\
    \Tilde{H}^{F}_{T}\lrrb{s',s,m,\alpha} = \int du\, u^3\, e^{-\sqrt{u^2 + z^2}}H^{F}_{T} (u,z,s,s',\alpha), 
\end{align}
where $u = p/T$ and $z = m/T$.

The functions inside the integrals can be obtained using analytical techniques. Define 
\begin{align}
    h(s) \equiv \tau(s)\,.
\end{align}For notational simplicity, we define the following variables
\begin{align}
    \frac{h(s')}{h(s)}   &\equiv \Tbh(s,s')  \equiv \Tbh \,,\\
    \frac{\alpha \times (s-s')}{h(s')} &\equiv g(s,s') \equiv g,
\end{align}
and
\begin{align}
     N^2 &\equiv \frac{1}{(1 - \Tbh^2)}\lrb{ \frac{u^2  + \Tbh^2g^2    + z^2}{ u^2} + \frac{g^2\Tbh^4}{u^2(1 - \Tbh^2)}  } \,,\\
    G   &\equiv \frac{g\Tbh^2}{u(1 - \Tbh^2) } \,.
\end{align}

Using the above definitions we can write
\begin{widetext}
   {\small  \begin{align*}
    H^{F} (u,z,s,s',\alpha) &=  - \Tbh \sqrt{(1-\Tbh^2)}N^2 \left\{ \frac{1}{2}\lrrb{ \sin^{-1}{ \frac{G  -  1}{ N }      } - \sin^{-1}{ \frac{\lrb{ G + 1}}{ N }      }   } \right.\nonumber \\
    &\left. + \frac{1}{2}\lrrb{  -\frac{1 + G}{N}\sqrt {1 - \lrb{\frac{1 + G}{N}}^2} + \frac{G-1}{N}\sqrt {1 - \lrb{\frac{G-1}{N}}^2 } } \right\},
\end{align*}

\begin{align*}
  H_{L} (u,z,s,s',\alpha) &= -\frac{\Tbh^3}{\sqrt{1-\Tbh^2}}\left\{   \lrb{\frac{ N^2}{2} + (g +G)^2 }\lrrb{ \sin^{-1}{\frac{G - 1}{N}} - \sin^{-1}{\frac{1 + G}{N}}  }   \right. \nonumber\\
                            &+ \frac{N^2}{2}\lrrb{\frac{1 + G}{N}\sqrt {1 - \lrb{\frac{1 + G}{N}}^2} - \frac{G-1}{N}\sqrt {1 - \lrb{\frac{G-1}{N}}^2 }} \nonumber\\
                            &-\left. 2N(g+G)\lrrb{\sqrt {1 - \lrb{\frac{1 + G}{N}}^2} - \sqrt {1 -\lrb{\frac{1 - G}{N}}^2} } \right\},
\end{align*}

\begin{align}
    H_{T} (u,z,s,s',\alpha) &= -\frac{\Tbh}{\sqrt{(1-\Tbh^2)}}\left\{ \lrb{ 1 -G^2 - \frac{ N^2}{2} }\lrrb{ \sin^{-1}{\frac{G - 1}{N}} - \sin^{-1}{\frac{1 + G}{N}}  }  \right. \nonumber\\
     &\left.  -\frac{N^2}{2}\lrrb{\frac{1 + G}{N}\sqrt {1 - \lrb{\frac{1 + G}{N}}^2} - \frac{G-1}{N}\sqrt {1 - \lrb{\frac{G-1}{N}}^2 }} \right. \nonumber \\
     &\left.  + 2NG\lrrb{\sqrt {1 - \lrb{\frac{1 + G}{N}}^2} - \sqrt {1 -\lrb{\frac{1 - G}{N}}^2} }   \right\}.
\end{align}}
\end{widetext}

\section{Iterative Solution Algorithm}
\label{sec:appIter}
Consider the integral equation
\begin{eqnarray}
    y(t) &=& K(t,t_0,y) v(t,t_0) \\
    &&+ \int_{t_0}^{t} dt' K(t,t',y)\frac{H(t,t',y(t'))}{\tau_R(y(t'))} f(y(t'))
\end{eqnarray}

where,
\begin{eqnarray}
   K(t,t',y) &=& \exp{\int_{t'}^{t}g(y(t''))dt''}   
\end{eqnarray}

and we are given the initial condition
\begin{equation}
    y(t_0) = y_0.
\end{equation}

This implies
\begin{equation}
    y(t_0) = v(t_0,t_0)
\end{equation}
as
\begin{equation}
    K(t,t,y) = 1
\end{equation}

The usual method for solving this equation is to consider an arbitrary functional form for the solution $y(t)$ and substitute it in the RHS and get a new solution. This process is continued until the iteration converges to some desired level of accuracy. The drawback of this method is that the speed of convergence depends on the initial guess $y(t)$. Here we take an alternate approach. If the function $y(t)$ is continuous then we approximate at each step $t_n = t_{n-1} + \Delta t_{n}$,
\begin{equation}
    y(t_{n-1}+\Delta t_{n}) = y(t_{n-1}) + \mathcal{O}(\Delta t_n)\,.
\end{equation}

Therefore with the initial condition $y(t_0) = v(t_0,t_0)$, we start with
\begin{equation}
    y(t_0 + \Delta t) \sim y(t_0 )
\end{equation}
and continue this procedure in order for each $n$ successively. We show elsewhere that after one iteration this produces an approximation for $y(t)$ accurate up to $\mathcal{O}(\Delta t^2)$. This process gets rid of the need for a good initial guess. It not only converges faster but also allows us to estimate order or error for our computation.

 \section{Inital Conditions}\label{App:InitCond}

 For the conformal case, we fix the anisotropy and normalization the same as the non-conformal case but vary the value of $\Lambda_0$ to match the initial energy density. The initial conditions for the distribution function were chosen to match \cite{Jaiswal:2022mdk} for comparison and are reproduced in the table \eqref{tab:Init_Conf}.

 \begin{widetext} 
\begin{table}[p]
    \begin{tabular}{|c|c|c|c|c|c|c|c|}\hline
      No.&  0 & 1 & 2 & 3 & 4 & 5 & 6\\ \hline
      $\Lambda_0$&  321.74 & 314 & 275 & 1089 & 198 & 500 & 500\\ \hline
      $\xi_0$  &  -0.832 & -0.908 & -0.949 & 1208.05 & -0.987 & 0  & 0\\\hline
\end{tabular}
\caption{Various values of $\Lambda_0$,$N_0$ and $\xi_0$ corresponding to different initial values of $(\pi/P)_0$ for m=0.}
\label{tab:Init_Conf}
\end{table} 
%\end{center}
\begin{table}[p]
    \begin{tabular}{|c|c|c|c|c|c|c|c|}\hline
       No.&  0 & 1 & 2 & 3 & 4 & 5 & 6\\ \hline
      $(\Pi/P)_{0}$&  0 & -0.25 & -0.37 & 0 & 0 & 0.25 & -0.85\\ \hline
      $(\pi/P)_{0}$ &  -1& -1 & -1& 0.99& -1.8& 0 & 0 \\ \hline
\end{tabular} 
\caption{Various initial values of $(\Pi/P)_{0}$,$(\pi/P)_{0}$ and their number code.}
\label{tab:Init_Pi_Jaiswal}
\end{table}
\begin{table}[p]
    \begin{tabular}{|c|c|c|c|c|c|c|c|}\hline
       No.&  0 & 1 & 2 & 3 & 4 & 5 & 6\\ \hline
      $m/\Lambda_0$&  0.016 & 4.808 & 10.89 & 0.294 & 1.818 & 2.023 & 20\\ \hline
      $N_0$  &  0.655 & $4 \times 10^{-5}$ & $2.5 \times 10^{-8}$ & 0.078 & 0.0632 & $1.06 \times 10^{-3}$  & $1.48 \times 10^{-13}$\\ \hline
      $\xi_0$  &  -0.832 & -0.908 & -0.949 & 1208.05 & -0.987 & 0  & 0\\\hline
\end{tabular} 
\caption{Various values of $m/\Lambda_0$,$N_0$ and $\xi_0$ corresponding to different initial values of $(\Pi/P)_{0}$,$(\pi/P)_{0}$  given in \eqref{tab:Init_Pi_Jaiswal}.}
\label{tab:Init_Jaiswal}
\end{table} 
\end{widetext}

\clearpage

\bibliography{main.bib}

\begin{thebibliography}{10}

\bibitem{STAR:2005gfr}
John Adams et~al.
\newblock {Experimental and theoretical challenges in the search for the quark
  gluon plasma: The STAR Collaboration's critical assessment of the evidence
  from RHIC collisions}.
\newblock {\em Nucl. Phys. A}, 757:102--183, 2005.

\bibitem{PhysRevD.46.229}
Jean-Yves Ollitrault.
\newblock Anisotropy as a signature of transverse collective flow.
\newblock {\em Phys. Rev. D}, 46:229--245, Jul 1992.

\bibitem{PhysRevC.82.039903}
B.~Alver and G.~Roland.
\newblock Erratum: Collision-geometry fluctuations and triangular flow in
  heavy-ion collisions [phys. rev. c 81, 054905 (2010)].
\newblock {\em Phys. Rev. C}, 82:039903, Sep 2010.

\bibitem{Romatschke:2009im}
Paul Romatschke.
\newblock {New Developments in Relativistic Viscous Hydrodynamics}.
\newblock {\em Int. J. Mod. Phys. E}, 19:1--53, 2010.

\bibitem{Heinz:2013th}
Ulrich Heinz and Raimond Snellings.
\newblock {Collective flow and viscosity in relativistic heavy-ion collisions}.
\newblock {\em Ann. Rev. Nucl. Part. Sci.}, 63:123--151, 2013.

\bibitem{Dusling:2015gta}
Kevin Dusling, Wei Li, and Bj\"orn Schenke.
\newblock {Novel collective phenomena in high-energy proton\textendash{}proton
  and proton\textendash{}nucleus collisions}.
\newblock {\em Int. J. Mod. Phys. E}, 25(01):1630002, 2016.

\bibitem{Li:2017qvf}
Wei Li.
\newblock {Collective flow from AA, pA to pp collisions \textendash{} Toward a
  unified paradigm}.
\newblock {\em Nucl. Phys. A}, 967:59--66, 2017.

\bibitem{Heller:2015dha}
Michal~P. Heller and Michal Spalinski.
\newblock {Hydrodynamics Beyond the Gradient Expansion: Resurgence and
  Resummation}.
\newblock {\em Phys. Rev. Lett.}, 115(7):072501, 2015.

\bibitem{Romatschke:2017acs}
Paul Romatschke.
\newblock {Relativistic Hydrodynamic Attractors with Broken Symmetries:
  Non-Conformal and Non-Homogeneous}.
\newblock {\em JHEP}, 12:079, 2017.

\bibitem{Jaiswal:2019cju}
Sunil Jaiswal, Chandrodoy Chattopadhyay, Amaresh Jaiswal, Subrata Pal, and
  Ulrich Heinz.
\newblock {Exact solutions and attractors of higher-order viscous fluid
  dynamics for Bjorken flow}.
\newblock {\em Phys. Rev. C}, 100(3):034901, 2019.

\bibitem{Chattopadhyay:2021ive}
Chandrodoy Chattopadhyay, Sunil Jaiswal, Lipei Du, Ulrich Heinz, and Subrata
  Pal.
\newblock {Non-conformal attractor in boost-invariant plasmas}.
\newblock {\em Phys. Lett. B}, 824:136820, 2022.

\bibitem{Jaiswal:2022udf}
Sunil Jaiswal, Jean-Paul Blaizot, Rajeev~S. Bhalerao, Zenan Chen, Amaresh
  Jaiswal, and Li~Yan.
\newblock {From moments of the distribution function to hydrodynamics: The
  nonconformal case}.
\newblock {\em Phys. Rev. C}, 106(4):044912, 2022.

\bibitem{Jankowski:2023fdz}
Jakub Jankowski and Micha\l{} Spali\'nski.
\newblock {Hydrodynamic Attractors in Ultrarelativistic Nuclear Collisions}.
\newblock 3 2023.

\bibitem{Kamata:2022jrc}
Syo Kamata, Jakub Jankowski, and Mauricio Martinez.
\newblock {Novel features of attractors and transseries in non-conformal
  Bjorken flows}.
\newblock 5 2022.

\bibitem{Dash:2020zqx}
Ashutosh Dash and Victor Roy.
\newblock {Hydrodynamic attractors for Gubser flow}.
\newblock {\em Phys. Lett. B}, 806:135481, 2020.

\bibitem{Chattopadhyay:2019jqj}
Chandrodoy Chattopadhyay and Ulrich~W. Heinz.
\newblock {Hydrodynamics from free-streaming to thermalization and back again}.
\newblock {\em Phys. Lett. B}, 801:135158, 2020.

\bibitem{Giacalone:2019ldn}
Giuliano Giacalone, Aleksas Mazeliauskas, and S\"oren Schlichting.
\newblock {Hydrodynamic attractors, initial state energy and particle
  production in relativistic nuclear collisions}.
\newblock {\em Phys. Rev. Lett.}, 123(26):262301, 2019.

\bibitem{Blaizot:2019scw}
Jean-Paul Blaizot and Li~Yan.
\newblock {Emergence of hydrodynamical behavior in expanding ultra-relativistic
  plasmas}.
\newblock {\em Annals Phys.}, 412:167993, 2020.

\bibitem{Strickland:2018ayk}
M.~Strickland.
\newblock {The non-equilibrium attractor for kinetic theory in relaxation time
  approximation}.
\newblock {\em JHEP}, 12:128, 2018.

\bibitem{Blaizot:2017ucy}
Jean-Paul Blaizot and Li~Yan.
\newblock {Fluid dynamics of out of equilibrium boost invariant plasmas}.
\newblock {\em Phys. Lett. B}, 780:283--286, 2018.

\bibitem{Behtash:2017wqg}
Alireza Behtash, C.~N. Cruz-Camacho, and M.~Martinez.
\newblock {Far-from-equilibrium attractors and nonlinear dynamical systems
  approach to the Gubser flow}.
\newblock {\em Phys. Rev. D}, 97(4):044041, 2018.

\bibitem{Alalawi:2022pmg}
Huda Alalawi and Michael Strickland.
\newblock {Far-from-equilibrium attractors for massive kinetic theory in the
  relaxation time approximation}.
\newblock {\em JHEP}, 12:143, 2022.
\newblock [Erratum: JHEP 07, 217 (2023)].

\bibitem{Gursoy:2014aka}
Umut Gursoy, Dmitri Kharzeev, and Krishna Rajagopal.
\newblock {Magnetohydrodynamics, charged currents and directed flow in heavy
  ion collisions}.
\newblock {\em Phys. Rev. C}, 89(5):054905, 2014.

\bibitem{Voronyuk:2011jd}
V.~Voronyuk, V.~D. Toneev, W.~Cassing, E.~L. Bratkovskaya, V.~P. Konchakovski,
  and S.~A. Voloshin.
\newblock {(Electro-)Magnetic field evolution in relativistic heavy-ion
  collisions}.
\newblock {\em Phys. Rev. C}, 83:054911, 2011.

\bibitem{Deng:2012pc}
Wei-Tian Deng and Xu-Guang Huang.
\newblock {Event-by-event generation of electromagnetic fields in heavy-ion
  collisions}.
\newblock {\em Phys. Rev. C}, 85:044907, 2012.

\bibitem{Zhao:2019crj}
Xin-Li Zhao, Guo-Liang Ma, and Yu-Gang Ma.
\newblock {Impact of magnetic-field fluctuations on measurements of the chiral
  magnetic effect in collisions of isobaric nuclei}.
\newblock {\em Phys. Rev. C}, 99(3):034903, 2019.

\bibitem{Kharzeev:2007jp}
Dmitri~E. Kharzeev, Larry~D. McLerran, and Harmen~J. Warringa.
\newblock {The Effects of topological charge change in heavy ion collisions:
  'Event by event P and CP violation'}.
\newblock {\em Nucl. Phys. A}, 803:227--253, 2008.

\bibitem{Denicol:2018rbw}
Gabriel~S. Denicol, Xu-Guang Huang, Etele Moln\'ar, Gustavo~M. Monteiro, Harri
  Niemi, Jorge Noronha, Dirk~H. Rischke, and Qun Wang.
\newblock {Nonresistive dissipative magnetohydrodynamics from the Boltzmann
  equation in the 14-moment approximation}.
\newblock {\em Phys. Rev. D}, 98(7):076009, 2018.

\bibitem{Denicol:2019iyh}
Gabriel~S. Denicol, Etele Moln\'ar, Harri Niemi, and Dirk~H. Rischke.
\newblock {Resistive dissipative magnetohydrodynamics from the Boltzmann-Vlasov
  equation}.
\newblock {\em Phys. Rev. D}, 99(5):056017, 2019.

\bibitem{Panda:2020zhr}
Ankit~Kumar Panda, Ashutosh Dash, Rajesh Biswas, and Victor Roy.
\newblock {Relativistic non-resistive viscous magnetohydrodynamics from the
  kinetic theory: a relaxation time approach}.
\newblock {\em JHEP}, 03:216, 2021.

\bibitem{Panda:2021pvq}
Ankit~Kumar Panda, Ashutosh Dash, Rajesh Biswas, and Victor Roy.
\newblock {Relativistic resistive dissipative magnetohydrodynamics from the
  relaxation time approximation}.
\newblock {\em Phys. Rev. D}, 104(5):054004, 2021.

\bibitem{Dash:2023kvr}
Ashutosh Dash and Ankit~Kumar Panda.
\newblock {Charged participants and their electromagnetic fields in an
  expanding fluid}.
\newblock 4 2023.

\bibitem{Alam:2021hje}
Sk~Noor Alam, Victor Roy, Shakeel Ahmad, and Subhasis Chattopadhyay.
\newblock {Electromagnetic field fluctuation and its correlation with the
  participant plane in Au+Au and isobaric collisions at sNN=200\,\,GeV}.
\newblock {\em Phys. Rev. D}, 104(11):114031, 2021.

\bibitem{Panda:2023akn}
Ankit~Kumar Panda, Reghukrishnan Gangadharan, and Victor Roy.
\newblock {Investigating the Role of Electric Fields on Flow Harmonics in
  Heavy-Ion Collisions}.
\newblock 1 2023.

\bibitem{STAR:2023jdd}
{Observation of the electromagnetic field effect via charge-dependent directed
  flow in heavy-ion collisions at the Relativistic Heavy Ion Collider}.
\newblock 4 2023.

\bibitem{Das:2022lqh}
Santosh~K. Das et~al.
\newblock {Dynamics of Hot QCD Matter -- Current Status and Developments}.
\newblock {\em Int. J. Mod. Phys. E}, 31:12, 2022.

\bibitem{Cercignani:2002rbe}
Carlo Cercignani and Gilberto~M. Kremer.
\newblock {\em {The Relativistic Boltzmann Equation: Theory and Applications}}.
\newblock Birkhäuser Basel, 2002.

\bibitem{Baym:1984np}
G.~Baym.
\newblock {Thermal equilibrium in relativistic heavy ion collisions}.
\newblock {\em Phys. Lett. B}, 138:18--22, 1984.

\bibitem{Florkowski:2013lza}
Wojciech Florkowski, Radoslaw Ryblewski, and Michael Strickland.
\newblock {Anisotropic Hydrodynamics for Rapidly Expanding Systems}.
\newblock {\em Nucl. Phys. A}, 916:249--259, 2013.

\bibitem{Florkowski:2013lya}
Wojciech Florkowski, Radoslaw Ryblewski, and Michael Strickland.
\newblock {Testing viscous and anisotropic hydrodynamics in an exactly solvable
  case}.
\newblock {\em Phys. Rev. C}, 88:024903, 2013.

\bibitem{Florkowski:2014sfa}
Wojciech Florkowski, Ewa Maksymiuk, Radoslaw Ryblewski, and Michael Strickland.
\newblock {Exact solution of the (0+1)-dimensional Boltzmann equation for a
  massive gas}.
\newblock {\em Phys. Rev. C}, 89(5):054908, 2014.

\bibitem{Romatschke:2003ms}
Paul Romatschke and Michael Strickland.
\newblock {Collective modes of an anisotropic quark gluon plasma}.
\newblock {\em Phys. Rev. D}, 68:036004, 2003.

\bibitem{Jaiswal:2022mdk}
Sunil Jaiswal, Subrata Pal, Chandrodoy Chattopadhyay, Lipei Du, and Ulrich
  Heinz.
\newblock {Far-from-equilibrium Attractor in Non-conformal Plasmas}.
\newblock {\em Acta Phys. Polon. Supp.}, 16(1):119, 2023.

\bibitem{Mitra:2022uhv}
Toshali Mitra, Sukrut Mondkar, Ayan Mukhopadhyay, Anton Rebhan, and Alexander
  Soloviev.
\newblock {Hydrodynamization in hybrid Bjorken flow attractors}.
\newblock 11 2022.

\bibitem{Florkowski:2017jnz}
Wojciech Florkowski, Ewa Maksymiuk, and Radoslaw Ryblewski.
\newblock {Coupled kinetic equations for fermions and bosons in the
  relaxation-time approximation}.
\newblock {\em Phys. Rev. C}, 97(2):024915, 2018.

\bibitem{Denicol:2019lio}
Gabriel~S. Denicol and Jorge Noronha.
\newblock {Exact hydrodynamic attractor of an ultrarelativistic gas of hard
  spheres}.
\newblock {\em Phys. Rev. Lett.}, 124(15):152301, 2020.

\bibitem{Dash:2022agb}
Ashutosh Dash and Victor Roy.
\newblock {Far-from-Equilibrium Hydrodynamic Attractor for~an~Azimuthally
  Symmetric System}.
\newblock {\em Springer Proc. Phys.}, 277:339--342, 2022.

\end{thebibliography}

\end{document}